# Fully solid-state all-2D micro-supercapacitors for on-chip energy storage

Bar Favelukis[1], Andreas Furchner[2], Barak Ratzker[1,3], Yiftach Kushnir[1], Asaf Nitsan[1], Rebeca Miyar[3], Yugal Rathod[4,5], Alexander Upcher[6], Nitzan Maman[6], Pini Shekhter[7], Varun Natu[4,5], Maxim Sokol[1,*]

[1] Department of Materials Science and Engineering, Tel Aviv University, P.O.B 39040, Ramat Aviv 6997801, Israel

[2] Helmholtz-Zentrum Berlin für Materialien und Energie GmbH, Nanoscale Solid–Liquid Interfaces, Schwarzschildstraße 8, 12489 Berlin, Germany

[3] Max Planck Institute for Sustainable Materials, Max-Planck-Str. 1, 40237 Düsseldorf, Germany

[4] Physical and Materials Chemistry Division, CSIR-National Chemical Laboratory India, Pune, 411008, India.

[5] Academy of Scientific and Innovative Research (AcSIR) Ghaziabad, UP 201002, India

[6] Ilse Katz Institute for Nanoscale Science and Technology, Ben-Gurion University of the Negev, P.O.B 653 Beer-Sheva 8410501, Israel

[7] Tel Aviv University Center for Nanoscience and Nanotechnology, Tel Aviv University, P.O.B 39040, Ramat Aviv 6997801, Israel

## Abstract

Integrating energy storage directly into microelectronic platforms remains a major challenge. Although solid-state micro-supercapacitors (MSCs) have been previously demonstrated, most still rely on liquid or solvent-containing gel electrolytes, which require additional encapsulation and complicate device integration. Here, we demonstrate wafer-scale fabrication of all-2D, fully solid-state MSCs by combining conductive $Ti_3C_2T_z$ MXene electrodes with a natural vermiculite (VMT) clay solid electrolyte. Both materials are solution-processable and can be independently patterned using microelectronics compatible fabrication techniques, allowing for their integration across 4-inch $Si/SiO_2$ wafers. The resulting devices showcase an active stack of only 28 nm thick, with an areal capacitance of 205 μF/cm$^2$ and volumetric energy density of ~15.5 mWh/cm$^3$. Notably, this work introduces a patterned 2D clay as a solid electrolyte in an on-chip electrochemical energy-storage device, extending the role of layered materials beyond electrodes to ion-conducting components. To investigate device behavior and degradation, we employ non-destructive imaging ellipsometry to map individual layer thicknesses and extract the spatial distribution of MXene conductivity across the chip. This approach reveals localized changes in electrode properties following prolonged electrochemical cycling, providing insight into degradation processes that are difficult to capture using conventional spatially averaged measurements. Together, these results establish a route towards fully solid-state, all-2D energy-storage devices and highlight the potential of solution-processable layered materials for integration into microelectronic systems.

## Introduction

The integration of energy storage into microelectronic platforms remains a key challenge for the development of next-generation electronics, as conventional discrete energy-storage components increase device footprint, packaging complexity, and system-level losses[1–3]. Micro-supercapacitors (MSCs) and thin-film batteries offer promising approaches toward on-chip energy storage; however, their integration into microelectronic platforms remains limited by material incompatibilities, processing constraints, and limitations in their scalable microfabrication[4].

Two-dimensional (2D) materials are a particularly attractive platform for addressing these challenges, owing to their atomic-scale thickness, mechanical flexibility, and compatibility with solution-based processing[5–7]. Among them, MXenes have emerged as promising electrode materials owing to their high conductivity, tunable surface chemistry, and exceptional electrochemical charge-storage properties[8,9]. Although lithographic patterning of MXene thin films has been demonstrated, their integration into wafer-scale microelectronic devices while preserving material quality remains challenging[10,11]. Moreover, MXene-based MSCs frequently rely on liquid or solvent-containing gel electrolytes[10,12]. Therefore, their practical implementation requires effective and robust encapsulation to prevent electrolyte leakage, solvent evaporation and moisture exchange, while ensuring stable performance during extended operating periods and varying temperatures[13]. At the microscale, such encapsulation can substantially increase the overall device volume and mass, thereby reduce the effective volumetric and gravimetric densities while complicating monolithic integration with microelectronic circuitry[14,15].

Solid-state electrolytes are essential for developing compact and integrable electrochemical devices[16]. Among solution-processable inorganic materials, 2D clays offer a promising class of solid electrolytes due to their intrinsically layered structure and ability to form ultrathin films from colloidal dispersions. Vermiculite (VMT), an abundant natural aluminosilicate $(Mg,Fe,Al)_3(Al,Si)_4O_{10}(OH)_2\cdot 4H_2O$), is particularly attractive due to its lamellar structure, which facilitates ion transport through interlayer channels and remains compatible with solution processing[17,18]. Although exfoliated VMT nanosheets can be assembled into ultrathin ion-conducting films, their integration into microelectronic devices presents a distinct fabrication challenge. In particular, lithographically defining ultrathin 2D clay films and integrating them with prepatterned electrodes while preserving their structural integrity and functionality remain largely unexplored.

Here, we demonstrate an all-2D, fully solid-state MSC by integrating metallic $Ti_3C_2T_z$ MXene electrodes with a layered VMT solid electrolyte. Both materials are deposited from colloidal solutions and independently patterned using conventional microfabrication techniques, combining selective wet etching of MXene with lithographic lift-off of the 2D clay electrolyte. This sequential fabrication approach enables the integration of electronically and ionically functional 2D materials directly on $Si/SiO_2$ substrates, providing a route towards wafer-scale fabrication of fully solid-state, ultrathin energy-storage devices. The resulting device presents an exceptionally thin out-of-plane footprint comprising an active stack thickness of only ~28 nm.

Beyond fabrication, characterizing the individual functional layers within ultrathin 2D heterostructures presents an additional challenge, particularly when investigating device operation and degradation. Here, we employ non-destructive imaging ellipsometry to spatially resolve the thicknesses of both MXene and VMT layers and the model-extracted conductivity of the MXene electrodes at the chip scale. By comparing pristine and electrochemically cycled devices, this approach enables the identification of localized changes in electrode properties, revealing degradation-prone regions that would otherwise be difficult to distinguish using spatially averaged measurements. Together, these results highlight the potential of layer-selective optical characterization for investigating functional 2D heterostructures integrated into microelectronic devices.

### Bulk precursors to 2D colloidal solutions

$Ti_3C_2T_z$ MXene and VMT were prepared as stable colloidal dispersions of two-dimensional flakes, enabling solution-based processing and thin-film assembly[19,20]. Among solution-processable 2D materials, MXenes are particularly attractive due to their demonstrated scalability to kilogram-level production[21,22]. $Ti_3C_2T_z$ was synthesized via the minimally intensive layer delamination (MILD) etching route[23], yielding delaminated flakes. The successful removal of the Al layers from the parent $Ti_3AlC_2$ MAX phase is exhibited by the characteristic X-ray diffraction (XRD) shift of the (002) reflection towards lower diffraction angles compared to that of $Ti_3AlC_2$ (Fig. 1a, top), indicating an increased interlayer spacing (Fig. 1a, bottom)[24]. Furthermore, the XRD pattern shows no detectable secondary phases, confirming the phase purity of the $Ti_3C_2T_z$. The resulting free-standing MXene film displays the expected layered morphology (Fig. S1a).

VMT was exfoliated using an ion-exchange-assisted delamination process[18,25], producing stable dispersions of layered nanosheets with expanded basal spacing (Fig. 1e, bottom) relative to the natural mineral precursor (Fig. 1e, top). The lamellar structure of VMT (Fig. S1b), its abundance and low cost, makes it an attractive ion-conducting building block for solid-state electrochemical systems[17,18,26,27]. The resulting colloidal dispersions remained stable for months in aqueous media without noticeable degradation[28].

Scanning electron microscopy (SEM) revealed ultrathin flakes for both materials (Fig. 1b,f), with lateral dimensions ranging from a few hundred nanometers to several micrometers, consistent with dynamic light scattering (DLS) measurements showing average hydrodynamic diameters of ~400 nm for MXene and ~600 nm for VMT (Fig. S2). The homogenous dispersions exhibited distinct optical signatures[29,30], as reflected in their characteristic UV–Vis spectra (Fig. 1h). Thin films deposited from concentrated colloids (Fig. 1c,g) retained high optical transparency (Fig. 1d), consistent with the reported absorption of ~3% of visible light per MXene layer (Fig. 1d, inset, right)[31] and the wide-bandgap nature of VMT, which remains highly transparent across the visible range (Fig. 1d, inset, left)[29]. These optical and colloidal characteristics confirm that both materials can be processed from stable aqueous dispersions into ultrathin films suitable for subsequent device assembly.

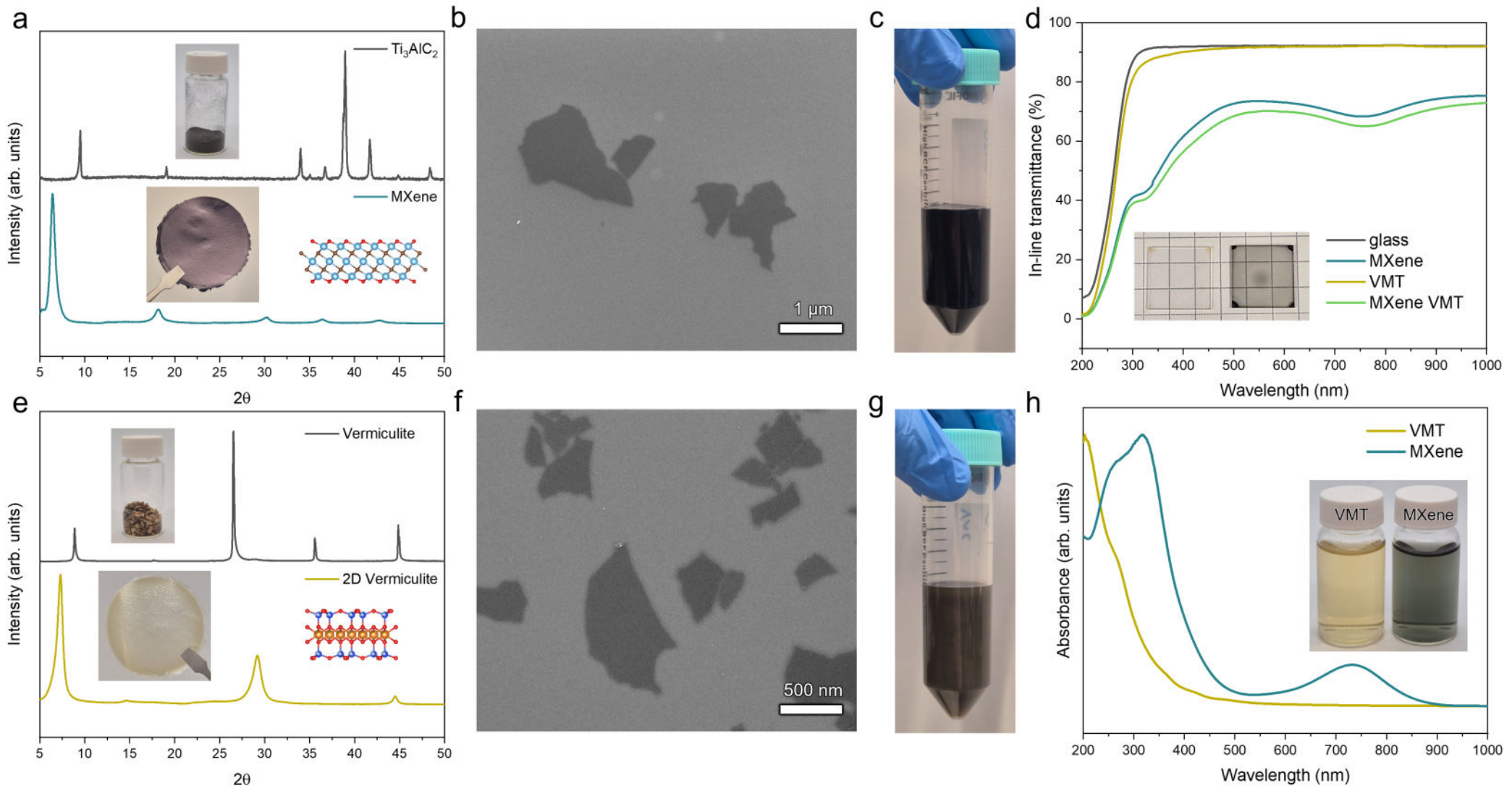

**Fig. 1. MXene and VMT characterization. a** XRD pattern of $Ti_3AlC_2$ powder (upper inset) and $Ti_3C_2T_z$ MXene freestanding film (bottom inset). **b** SEM image of MXene single flakes. **c** Photograph of the MXene colloidal solution. **d** UV–Vis transmittance spectra of MXene and VMT, inset shows glass substrates coated with MXene and VMT thin films. **e** XRD pattern of VMT particles (upper inset) and delaminated VMT freestanding film (bottom inset). **f** SEM image of VMT single flakes. **g** Photograph of the VMT colloidal solution. **h** UV–Vis spectra of diluted MXene and VMT solutions (inset: photographs of the solutions).

## Wafer-scale fabrication of on-chip all-2D solid-state micro-supercapacitor

The integrated MXene@VMT MSC fabrication process is summarized in Fig. 2. MXene electrodes were first prepared by spin coating $Ti_3C_2T_z$ onto $Si/SiO_2$ substrates, followed by a dilute HCl spin-cleaning step to remove residual halide byproducts, yielding continuous ultrathin films with a thickness of ~6.5 nm[32]. The electrode geometry was then defined by standard photolithography and selective wet etching of MXene using our previously developed process (Fig. 2a)[11]. This wet-etching route avoids direct plasma exposure, which can damage or oxidize ultrathin 2D materials edges, while preserving the spatial fidelity required for lithographically defined microscale electrodes[33–35]. The resulting interdigitated MXene electrodes showed uniform coverage across a full 4″ wafer (Fig. 2c), with well-defined comb-shaped architectures and minimal edge roughness confirmed by SEM (Fig. 2d and Fig. S3).

Integration of the VMT electrolyte requires a second lithographic step (Fig. 2b). During photolithographic processing, the exposed MXene regions become poorly wettable by the aqueous VMT colloid, preventing uniform electrolyte deposition. To overcome this limitation, a short UV–ozone treatment was applied to remove residual organic contaminants[36] and induce mild surface modification of the MXene[37]. This treatment restores surface wettability, enabling uniform deposition of the delaminated VMT nanosheets and sufficient adhesion to withstand the subsequent lift-off process involving sonication in hot DMF. To evaluate the effect of UV–ozone exposure on the underlying MXene electrodes, XPS measurements were performed (Fig. 2g). Ti 2p spectra reveal only minor oxidation after the 5 min treatment used for fabrication (full spectra are shown in Figs. S4–S7), whereas prolonged exposure (20 min) leads to pronounced oxidation (Fig. S8). Thus, the short UV–ozone treatment enables successful lithographic patterning of the ultrathin 2D clay electrolyte while limiting oxidation of the underlying MXene electrodes. Wafer-scale VMT integration is illustrated in Fig. 2e, while SEM confirms coverage of the patterned MXene@VMT MSCs (Fig. 2f).

To enable non-destructive, spatially resolved characterization of the integrated MXene@VMT devices at the chip scale, imaging ellipsometry was applied based on our previously developed methodology[38]. The distinct optical properties of MXene and VMT enable layer-selective mapping of the individual film thicknesses and the model-extracted effective conductivity of the MXene electrodes (Fig. 2h). The optically extracted MXene conductivity was approximately 9200 S $cm^{-1}$ at the non-intercalated pads and 7200 S $cm^{-1}$ at the intercalated electrodes. Additional high-resolution thickness and conductivity maps (Figs. S9–S14) demonstrate the spatial uniformity of the patterned 2D layers. Combined atomic force microscopy (AFM) and imaging ellipsometry measurements confirm an active stack thickness of ~28 nm, comprising ~8 nm MXene electrodes and ~20 nm VMT electrolyte films (Fig. 2h and S9–S15).

Together, the selective wet-etching and lift-off processes enable sequential patterning of electronically and ionically functional 2D materials across a 4″ wafer. By combining solution-based deposition with conventional photolithography, the fabrication process avoids vacuum deposition and high-temperature material growth, providing a scalable approach for integrating ultrathin all-2D electrochemical devices into planar microfabrication workflows[39,40].

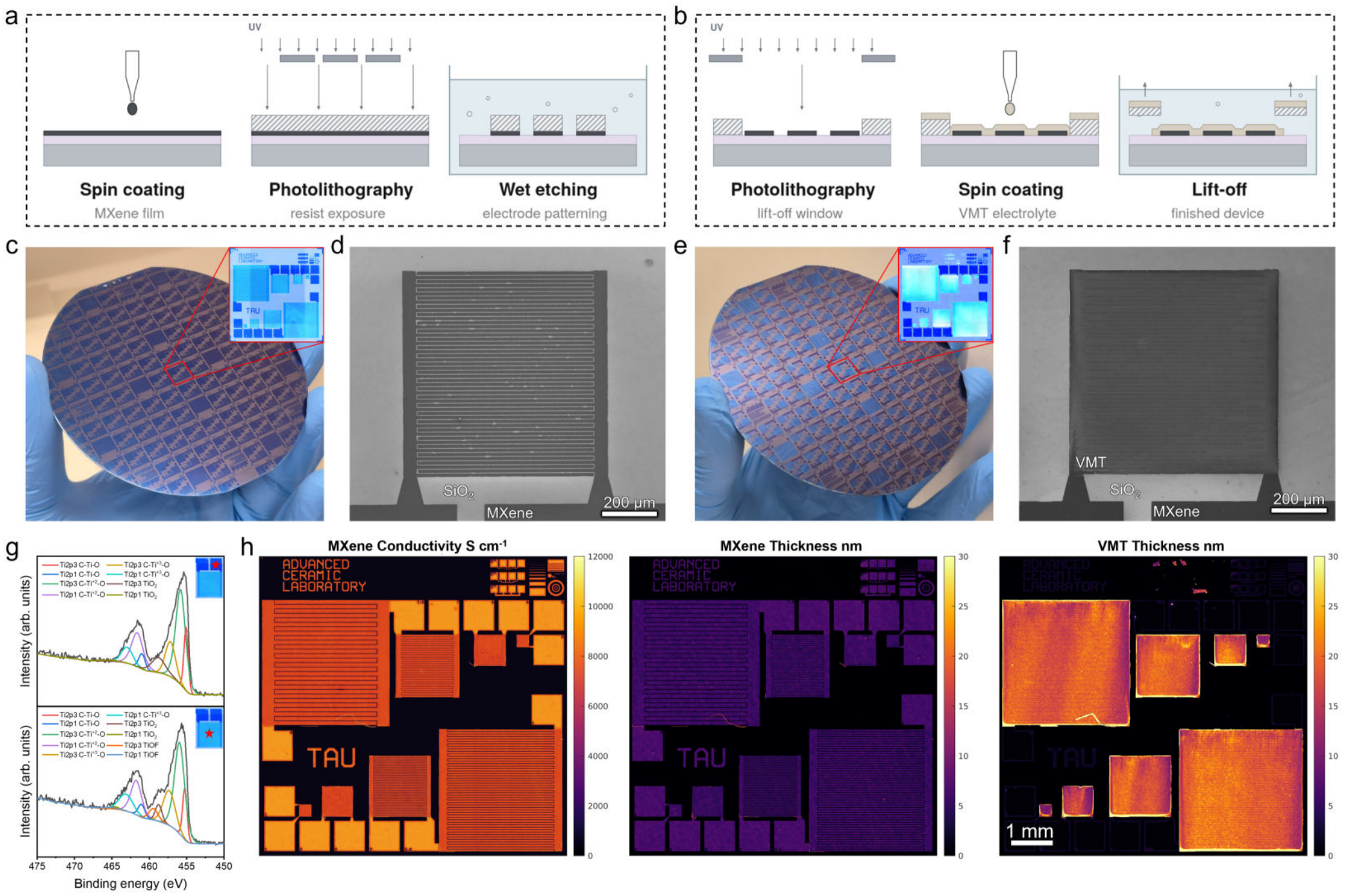

**Fig. 2. Fabrication process of wafer-scale all-2D solid-state on-chip micro-supercapacitor (MSC). a** Illustration of deposition and patterning steps of MXene electrodes. **b** Illustration of deposition and patterning steps of VMT electrolyte. **c** Photograph of a 4″ $SiO_2$ wafer with various MXene electrodes, inset shows optical microscope image of a chip unit. **d**

SEM image of MXene comb-shaped electrodes. **e** Photograph of a 4″ $SiO_2$ wafer with MXene electrodes and VMT electrolyte, inset shows optical microscope image of a chip unit. **f** SEM image of the integrated MXene@VMT MSC. **g** Ti 2p XPS spectra of the reference MXene pad and patterned MXene electrodes. **h** Imaging ellipsometry of the completed chip showing MXene conductivity, MXene thickness, and VMT thickness.

## Electrochemical performance of the MXene@VMT micro-supercapacitors

The electrochemical performance of the MXene@VMT MSCs is presented in Fig. 3. Cyclic voltammetry (CV) measurements recorded over scan rates of 1–2000 $mVs^{-1}$ (Fig. 3a,b) exhibit quasi-rectangular profiles at low scan rates, with increasing distortion at higher rates, consistent with capacitive charge storage limited by finite ion transport within the ultrathin solid-state architecture[3,41]. Details of the electrochemical performance calculations are provided in the Supplementary Information. The corresponding areal capacitance decreases from ~215 $\mu F\ cm^{-2}$ to ~10 $\mu F\ cm^{-2}$ with increasing scan rate (Fig. 3c), reflecting reduced ion accessibility at shorter timescales[42].

The electrochemical stability window was evaluated by extending the applied potential range. Long-term CV cycling measurements demonstrated stable operation at ±1 V and sustained cycling at ±1.25 V, retaining ~80% of the initial capacitance after 3000 cycles, whereas higher applied potentials (≥±1.5 V) resulted in rapid degradation (Fig. S16a–h). These measurements establish a practical bipolar cycling range of ±1–1.25 V, corresponding to a total potential sweep span of 2–2.5 V. Notably, this bipolar cycling capability is achieved within an ultrathin, fully solid-state architecture, in contrast to representative MSCs relying on liquid or gel-based electrolytes[3,12,43,44].

To analyze the failure mechanisms of the MXene@VMT devices, CV measurements over an extended bipolar voltage window of ±5 V (50 $mV\ s^{-1}$) were carried out, revealing irreversible degradation accompanied by visible micrometer-scale bubble formation (Fig. S17a,c), consistent with MXene oxidation and associated gas evolution[45,46]. Voltage-window-dependent CV measurements further showed a progressive decrease in capacitance at higher potentials (Fig. S17b), while post-cycling optical microscopy confirmed the formation of defects, bubbles beneath the VMT layer, and local electrode disconnection after severe overpolarization (Fig. S18a–d).

Galvanostatic charge–discharge (GCD) profiles recorded at different applied currents (Fig. 3d) exhibit a capacitive charge–discharge behavior with deviations from ideal linearity, reflecting nonideal charge-storage dynamics and resistive contributions[47,48]. The extracted areal capacitance decreases from ~249 $\mu F\ cm^{-2}$ to ~85 $\mu F\ cm^{-2}$ with increasing current (Fig. 3e), consistent with the current-density-dependent rate behavior shown in Fig. 3f and with the devices shown in Fig. S19–S20. Despite the active stack thickness of only ~28 nm, comprising ~8 nm MXene electrodes and a ~20 nm solid-state VMT electrolyte, the MXene@VMT MSCs achieve an areal energy density of up to ~0.0124 $\mu Wh\ cm^{-2}$, corresponding to a volumetric energy density of ~15.5 $mWh\ cm^{-3}$. At higher current densities, the device reaches an areal power density of up to ~1.09 $\mu W\ cm^{-2}$, corresponding to ~1.37 $W\ cm^{-3}$. A comprehensive comparison with representative state-of-the-art on-chip MSCs[3,12,43,44] is summarized in Table S1, highlighting the competitive volumetric energy storage performance achieved within an exceptionally thin, fully solid-state, all-2D architecture.

Open-circuit potential decay measurements (Fig. 3g) show a gradual voltage decrease from 1.25 V to 0.1 V over ~4.5 h, indicating moderate self-discharge in the nanoconfined electrode–electrolyte architecture[48,49]. Electrochemical impedance spectroscopy (EIS) measurements (Fig. 3h) show a small high-frequency semicircle followed by an inclined low-frequency response, consistent with capacitive charge storage accompanied by nonideal transport and resistive contributions[50]. Bias-dependent EIS measurements (Fig. S21a–c) reveal potential-dependent changes in the impedance spectra, indicating that the device response evolves with applied bias. Compared with the nearly vertical impedance response of bare MXene electrodes, the VMT-covered devices exhibit a pronounced semicircle and inclined low-frequency response, demonstrating the substantial influence of the solid electrolyte on the device impedance (Fig. S22).

GCD cycling at a fixed current density (Fig. 3i) demonstrates stable and reproducible charge–discharge behavior. Long-term cycling over ~900 cycles for multiple devices shows moderate capacitance decay (~35%), together with a minor increase in low-frequency impedance (Fig. S23a–f), indicating gradual performance degradation during extended operation. Temperature-dependent electrochemical measurements of the MXene@VMT devices heated under ambient conditions reveal a reversible decrease in capacitance from ~96 to ~7 μF $cm^{-2}$ between 18 and 85 °C, followed by full recovery upon cooling (Fig. S24a–f). In contrast, temperature-dependent impedance measurements of a free-standing VMT film in a Swagelok cell, under constant relative humidity across the investigated temperature range, show thermally activated ionic conductivity, with an apparent activation energy of ~0.29 eV (Fig. S25a,b). This thermally activated behavior is consistent with previous reports by Qian et al.[51] and Zhang et al.[52] on ion transport through hydrated 2D nanochannels. An independent in-plane four-terminal measurement of a free-standing VMT film yields a conductivity of $9.61 \cdot 10^{-4}$ S $cm^{-1}$ (Fig. S25c), consistent with efficient lateral ionic transport along the layered structure[18].

Together, these results suggest that the decrease in device capacitance is primarily associated with reversible changes in the hydration state and charge-storage environment of the nanoconfined VMT electrolyte, rather than a reduction in intrinsic ionic mobility, as elevated temperature can reduce or reorganize retained/interlayer water and modify ionic charge-storage pathways[27,51,52].

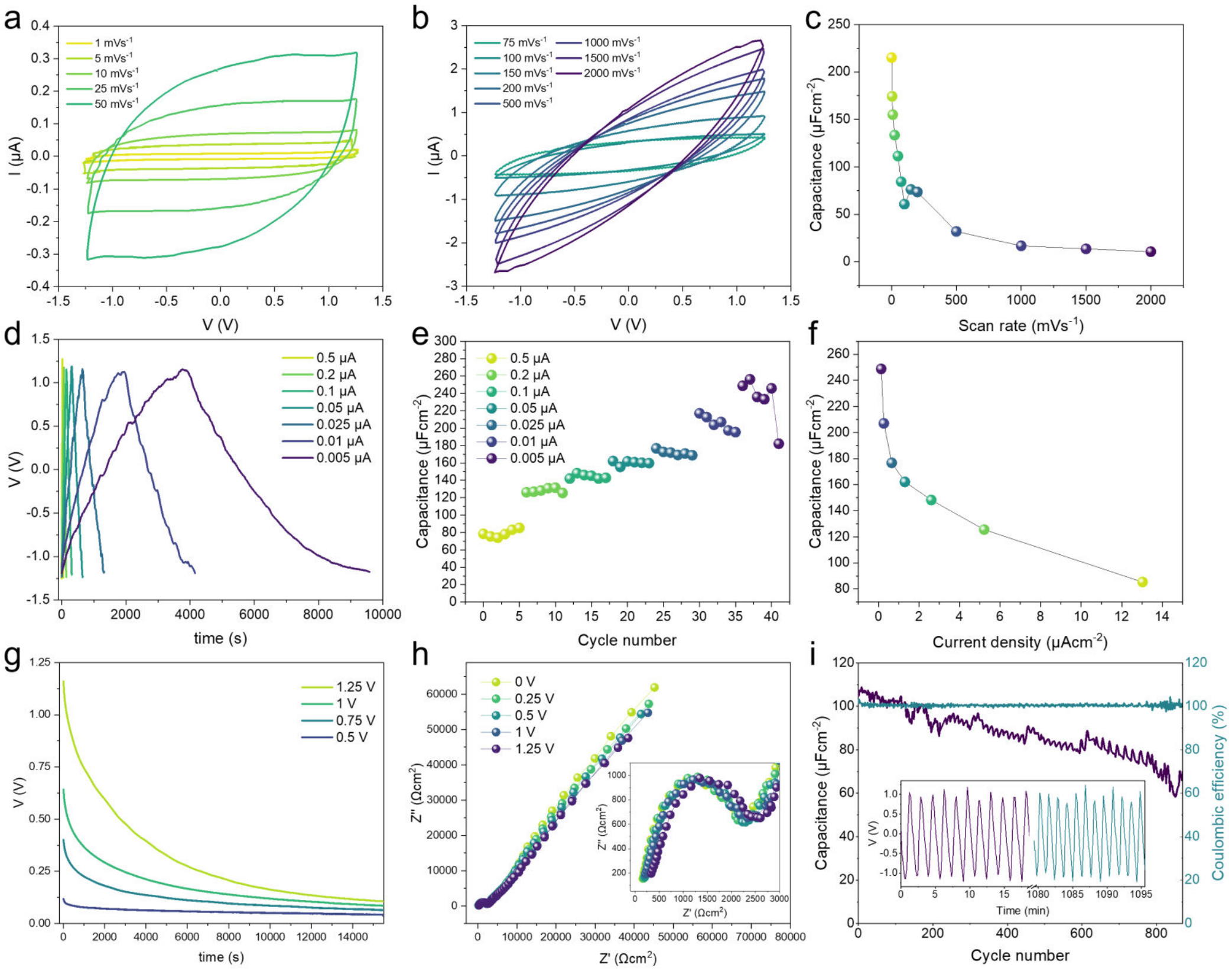

**Fig. 3. Electrochemical performance of the MXene@VMT MSC. a,b** CV curves of the MXene@VMT MSC measured at different scan rates. **c** Dependence of areal specific capacitance on scan rate. **d** Galvanostatic charge discharge (GCD) profiles of the MXene@VMT MSC at different currents. **e** Rate performance of the MXene@VMT MSC at different currents. **f** Dependence of areal specific capacitance on current density. **g** Decay of open-circuit potential of the MXene@VMT MSC after charging to different potentials. **h** EIS Nyquist plots recorded at different bias potentials. **i** Capacitance retention and Coulombic efficiency during galvanostatic cycling at a current density of 10 μA $cm^{-2}$. Inset: representative GCD profiles.

## Post-operation MXene@VMT thin-film characterization

MXene undergoes changes in optical response upon ion intercalation[53–56], allowing in situ optical microscopy to monitor electrode contrast during electrochemical cycling. To investigate the onset and spatial evolution of degradation under elevated potentials, in situ optical microscopy was performed while the device was cycled at 200 mV $s^{-1}$ under progressively increasing voltage windows from ±2 to ±5 V in 0.5 V increments (Video S1 and Fig. S17b). Reversible changes in electrode contrast were observed during cycling, consistent with ion intercalation and deintercalation processes in the MXene electrodes. At higher voltage windows, irreversible changes became evident: small bubbles first appeared near the base of the electrodes at ~±3 V and became more pronounced at higher voltages. Further overpolarization caused parts of the interdigitated electrodes to lose their reversible optical contrast, indicating localized loss of electrochemical activity. These observations are consistent with the progressive decrease in capacitive response at elevated voltage windows (Fig. S17b), as well as the optical defects and electrode disconnection observed after high-voltage stress and extended cycling (Figs. S18 and S26).

Imaging ellipsometry was further employed to investigate spatially resolved changes in the MXene@VMT devices following electrochemical operation. After 1000 charge–discharge cycles at 5 μA $cm^{-2}$, localized degradation was observed primarily near the electrode edges, where a pronounced decrease in the extracted MXene conductivity was accompanied by a decrease in the extracted MXene thickness and a corresponding increase in the extracted VMT thickness (Fig. 4a). Importantly, these thickness variations should not be interpreted as direct geometrical changes of the individual layers. Oxidation of Ti-based MXenes is known to strongly modify their dielectric function, including the refractive index (*n*) and extinction coefficient (*k*), while suppressing the metallic Drude response and shifting the optical response toward more $TiO_x$-like behavior[46,57,58]. In pristine devices, the distinct optical properties of MXene and VMT (Fig. 1d,h) enable their respective thicknesses and the MXene conductivity to be extracted using a multilayer optical model. However, in degraded regions, the optical response can no longer be fully described using the dielectric function of pristine $Ti_3C_2T_z$. Consequently, part of the altered MXene optical response may be assigned to the VMT layer, resulting in an apparent decrease in MXene thickness and a corresponding increase in VMT thickness. This effect was considerably more pronounced in devices intentionally degraded under high-voltage operation (50 mV $s^{-1}$, 5 V; Video S2 and Fig. S27), where the positively biased electrodes exhibited a spatially extended decrease in MXene thickness and conductivity, together with a substantial apparent increase in VMT thickness. The reduction in MXene conductivity is consistent with partial oxidation[59] and may be further enhanced by extensive ion intercalation[60,61]. Importantly, identifying the spatial distribution of these changes highlights regions particularly susceptible to degradation, providing valuable insight into potential degradation mechanisms and guidance for future electrode designs aimed at improving operational stability.

AFM measurements provided an independent assessment of the device topography and film thickness before and after electrochemical operation. Bare MXene electrodes exhibited an average thickness of ~8 nm, corresponding to approximately five intercalated MXene layers[38,62]. The patterned electrodes displayed well-defined, sharp step profiles, consistent with the high pattern fidelity achieved by selective wet etching[11]. Following VMT deposition, the total thickness of the MXene@VMT heterostructure reached ~28 nm in the central active region. A lift-off edge effect[10,32] was observed near the patterned boundaries, resulting in locally increased thicknesses of the edges. Importantly, AFM measurements performed on devices cycled within the practical operating window (up to ±1.25 V at 200 mV $s^{-1}$) and stopped at different states of charge revealed no significant changes in surface morphology or film thickness (Fig. 4b and Fig. S15).

Finally, cross-sectional STEM imaging provided direct structural characterization of the MXene@VMT heterostructure following electrochemical operation (Fig. 4c). The cycled device retained a well-defined multilayered architecture, with no clear evidence of interfacial delamination in the examined region. Together with the AFM results, these observations demonstrate the preservation of the overall device morphology under the investigated operating conditions, while imaging ellipsometry reveals localized changes in the optical and electronic properties of the MXene electrodes following prolonged cycling.

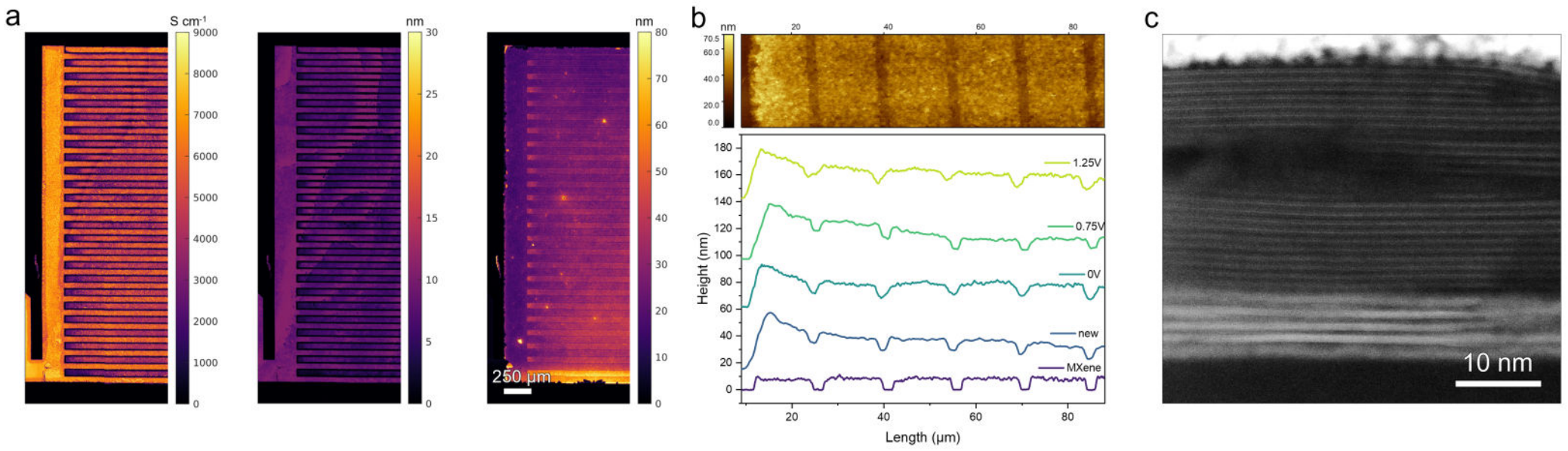

**Fig. 4. Post-operation characterization of MXene@VMT MSCs. a** Imaging ellipsometry of the device after 1K GCD cycling showing MXene conductivity, MXene thickness, and VMT thickness. **b** AFM analysis of the MXene@VMT device after cycling to different terminal potentials. **c** Cross-sectional STEM image of the MXene@VMT device. The region at the center of the VMT suffered from beam damage during examination.

## Conclusions

This work demonstrates the fabrication of all-2D, all-solid-state MSCs by integrating metallic $Ti_3C_2T_z$ MXene electrodes with a lithographically patterned VMT clay electrolyte. By combining selective wet etching of MXene with lift-off patterning of fully delaminated VMT nanosheets, we achieve independent spatial control over the electronic and ionic components using conventional microfabrication techniques. This approach enables wafer-scale fabrication on 4-inch $Si/SiO_2$ substrates, yielding an active device stack only ~28 nm thick. Despite their ultrathin architecture, the MSCs achieve an areal capacitance of 160 µF $cm^{-2}$ and a volumetric energy density of approximately 275 mWh $cm^{-3}$, normalized to the active electrode volume.

Non-destructive imaging micro-ellipsometry enables spatially resolved, layer-selective characterization of the integrated MXene@VMT heterostructure, providing simultaneous maps of the individual MXene and VMT film thicknesses and the model-extracted MXene conductivity. By comparing pristine and cycled devices, this approach reveals localized conductivity losses, primarily near the electrode edges, accompanied by changes in the apparent thicknesses of the constituent layers. Importantly, identifying the spatial distribution of progressive electrode degradation provides valuable insight into potential degradation mechanisms and highlights vulnerable regions of the device architecture, offering guidance for future improvements in electrode design and operational stability.

More broadly, this work demonstrates how conventional microfabrication techniques can be adapted to integrate distinct classes of solution-processed 2D materials into functional, ultrathin solid-state devices. Combining independent lithographic patterning with layer-selective, non-destructive characterization provides opportunities to investigate and optimize electronic–ionic interactions within integrated 2D heterostructures. Beyond MSCs, this approach may enable a broader range of solid-state microelectronic devices featuring coupled electronic and ionic transport.

## Methods

### MAX Phase Synthesis

Titanium carbide (TiC, 99.5%, Alfa Aesar), titanium (Ti, 99.7%, Strem), and aluminum (Al, 99.7%, Strem) powders were mixed in a 2:1:1.1 ratio and tumbler-mixed with stainless steel balls at 200 rpm for 18 h. The blended powders were placed in an alumina crucible and heat-treated under a 5% $H_2$/Ar atmosphere in a tube furnace at 1450 °C for 3 h, with a heating/cooling rate of 5 °C/min. The obtained $Ti_3AlC_2$ MAX phase was ground and sieved to obtain a powder with a particle size of ≤ 20 μm.

### MXene Synthesis

$Ti_3C_2T_z$ MXene was synthesized by selectively etching $Ti_3AlC_2$. One gram of MAX powder was slowly added to a solution of 20 mL 10.2 M HCl (32%, Bio-Lab) and 1.6 g lithium fluoride (LiF, 99%, Strem) at 45 °C and stirred for 24 h. After etching, the mixture was centrifuged at 3500 rpm several times with DI water until the MXene exhibited osmotic swelling. The solution was sonicated in an ultrasonic bath for 30 min while being maintained at 5 °C. After sonication, the suspension was centrifuged at 3500 rpm for 30 min to separate any unexfoliated sediment. The MXene colloidal solution was bubbled with $N_2$ for 15 min and stored at 4 °C. A 5 mL aliquot of the colloid was dried and weighed to determine the MXene concentration, which was approximately 12–14 g/L.

### Delaminated Vermiculite synthesis

A 2D delaminated VMT $(Mg,Fe,Al)_3(Al,Si)_4O_{10}(OH)_2{\cdot}4H_2O$) colloidal solution was synthesized via an ion-exchange method[18]. Three grams of vermiculite particles (Vermiculite 2-3 mm, Sigma-Aldrich) were immersed in 200 mL of saturated sodium chloride (NaCl, 99%, Strem) solution (72 g) and refluxed with stirring for 24 h. The mixture was centrifuged at 3500 rpm several times with DI water and then dispersed in 100 mL of a 2 M lithium chloride (LiCl, 99%, Strem) solution (8.48 g), followed by refluxing for another 24 h. The mixture was washed until osmotic swelling occurred, then sonicated in an ultrasonic bath for 1 h while being maintained at 5 °C. After sonication, the suspension was centrifuged at 3500 rpm for 30 min to separate any unexfoliated sediment. A 3 mL aliquot of the colloid was dried and weighed to determine the VMT concentration, which was approximately 21–23 g/L.

### MXene Thin-Film Deposition and patterning

MXene deposition and patterning were performed using the protocols showcased in our previous studies[11,32]. $SiO_2$ (4”, n-type, 100 nm wet thermal oxide, UniversityWafer, Inc.) wafers were piranha-treated using a 1:3 $H_2O_2/H_2SO_4$ solution ($H_2SO_4$ 98%, SDFCL; $H_2O_2$ 35%, Thermo Fisher Scientific) to increase hydrophilicity, then spin-coated with MXene solution (11 g/L) at 2000 rpm for 1 min, followed by spin-cleaning with 0.5 M HCl at 2000 rpm for 1 min.

The MXene-coated wafers were heated to 200 °C for 5 min to remove residual intercalated water, then spin-coated with HMDS (99%, Sigma-Aldrich) at 3000 rpm for 40 s and heated again to 200 °C for 5 min. The MXene electrodes were patterned by photolithography; the substrate was spin-coated with Futurrex NR-1500PY photoresist at 3000 rpm for 40 s, baked at 150 °C for 1 min, exposed with 285 mJ $cm^{-2}$ using a mask aligner (SUSS MA6), postbaked at 100 °C for 1 min, and developed in RD6 solution for 17 s. MXene removal by wet etching was performed using a 1:3 $H_2O_2$/HCl solution with 0.5 vol % Triton X-100 for 1 min. After etching, the photoresist was removed using acetone, IPA, and water.

**VMT Thin-Film Deposition and patterning**

VMT electrolyte was patterned by photolithography. The MXene-patterned wafers were spin-coated with HMDS at 4000 rpm for 40 s, then spin-coated with AZ1518 photoresist at 4000 rpm for 40 s, baked at 110 °C for 1.5 min, exposed with 155 mJ $cm^{-2}$ using a mask aligner, and developed in AZ-726 solution for 40 s. For process calibration, some of the wafers were diced using a DISCO DAD300 dicing saw. The developed wafers and dies were UV ozone treated for 5 min to increase hydrophilicity, then spin-coated with VMT solution (21 g/L) at 2000 rpm for 1 min, followed by heating at 150 °C for 15 s to remove residual intercalated water. The VMT-covered photoresist was removed using lift-off in an ultrasonic bath for 5 s in 100 °C DMF, followed by rinsing with acetone and IPA.

**Characterization**

XRD measurements were performed over a 2θ range of 5–85° with 0.02° steps (Malvern Panalytical Aeris diffractometer). XPS analysis was carried out using an ESCALAB QXi (Thermo Scientific). The morphology of the MXene and VMT flakes was examined via high-resolution SEM (ZEISS Gemini 300). UV–Vis in-line transmittance and extinction spectra of MXene and VMT were collected from spin-coated glass substrates (2000 rpm) and diluted colloidal solutions in quartz cuvettes (Shimadzu UV-2600 spectrophotometer), operated in transmission mode over the 200–800 nm wavelength range. Hydrodynamic size distribution of MXene and VMT flakes were analyzed through DLS analysis using a Zetasizer Nano ZS (Malvern Instruments, Ltd.). Deposition quality and MXene@VMT devices were examined using SEM (JEOL JSM-IT200). Device thickness was measured using an NX12 AFM (Park Systems) equipped with a PPP-NCHR probe (nominal resonance frequency: 330 kHz; spring constant: 42 N/m). All scans were acquired in non-contact mode. Confocal microscope analysis of MXene@VMT was performed using a Lext (OLS5100 LEXT, Olympus). Device-scale images of thickness and electrical conductivity of the MXene electrodes, as well as the thickness of the VMT electrolyte layer, were obtained noninvasively through micro-imaging ellipsometry (Accurion EP4, Park Systems GmbH) following our previously reported methodology.[38]. STEM imaging of the MXene@VMT cross-section was performed using a probe Cs-corrected (S-CORR) Thermo Fisher Scientific Spectra 200 (S)TEM operating at 200 kV. The lamellae for STEM analysis were prepared using a dual-beam focused ion beam (FIB; FEI Helios G4 UC).

Electrochemical characterization of MXene@VMT devices: All electrochemical measurements were conducted using a BioLogic VSP-300 potentiostat in a two-electrode configuration. The devices were contacted using an Ossila micromanipulator with tungsten probes directly attached to the MXene pads.

Areal capacitance was determined from the integrated CV loop area and the discharge charge recorded during GCD measurements, normalized to the area of a single MXene electrode. Energy and power densities were calculated by integrating the experimental GCD profiles over the positive-polarity discharge interval (+1.25 to 0 V), with the negative-polarity response evaluated separately. Areal and volumetric energy and power densities were normalized to the combined area and volume of both MXene electrodes, respectively. Detailed equations and normalization procedures are provided in the Supporting Information.

Out-of-plane ionic conductivity was measured using a free-standing VMT film placed between two glassy carbon electrodes in a Swagelok cell. Temperature-dependent EIS measurements were performed under controlled relative humidity. The activation energy was determined from the Arrhenius dependence of ionic conductivity. In-plane conductivity of free-standing VMT film was measured using a custom-built four-terminal setup consisting of four parallel Pt wires mounted on a PMMA support.

**Supporting information**

Additional characterization data of MXene and VMT films and device performances (PDF).

**Acknowledgements**

The authors are grateful to Erez Benjamin and Nirit Porecki Shamay for their assistance with photolithography, and to Inna Shekhtman for her assistance with photomask fabrication. All are from the Center for Nanoscience and Nanotechnology at Tel Aviv University.

This work was supported by the Israel Science Foundation (grant No. 3407/26).

A. Furchner acknowledge support from the European Regional Development Fund (EFRE 1.8/13).

**Data availability**

The data supporting the findings of this study are available within the Article and its Supplementary Information. Additional data are available from the corresponding author upon reasonable request.

## Author contributions

**B.F.** conceived the study, developed and fabricated the devices, performed the majority of the experiments and characterization, analyzed the data, prepared the figures, and wrote the original manuscript. **A.F.** performed the microellipsometry measurements and analysis. **B.R.** contributed to manuscript writing, visualization and editing. **Y.K.** performed the AFM characterization. **A.N.** performed the optical microscopy characterization. **R.M.** assisted in electrochemical data analysis and writing and reviewing. **Y.R.** contributed to electrochemical measurements. **A.U.** performed the STEM characterization. **N.M.** performed the FIB preparation. **P.S.** performed the XPS measurements and analysis. **V.N.** provided guidance on electrochemical measurements, analysis, and interpretation. **M.S.** supervised the study, contributed to its conceptualization and interpretation, and revised the manuscript. All authors discussed the results, reviewed and approved the final manuscript.

# Fully solid-state all-2D micro-supercapacitors for on-chip energy storage

## Supplementary information

Bar Favelukis[1], Andreas Furchner[2], Barak Ratzker[1,3], Yiftach Kushnir[1], Asaf Nitsan[1], Rebeca Miyar[3] Yugal Rathod[4,5], Alexander Upcher[6], Nitzan Maman[6], Pini Shekhter[7], Varun Natu[4,5], Maxim Sokol[1,*]

[1] Department of Materials Science and Engineering, Tel Aviv University, P.O.B 39040, Ramat Aviv 6997801, Israel

[2] Helmholtz-Zentrum Berlin für Materialien und Energie GmbH, Nanoscale Solid–Liquid Interfaces, Schwarzschildstraße 8, 12489 Berlin, Germany

[3] Max Planck Institute for Sustainable Materials, Max-Planck-Str. 1, 40237 Düsseldorf, Germany

[4] Physical and Materials Chemistry Division, CSIR-National Chemical Laboratory India, Pune, 411008, India.

[5]Academy of Scientific and Innovative Research (AcSIR) Ghaziabad, UP 201002, India

[6] Ilse Katz Institute for Nanoscale Science and Technology, Ben-Gurion University of the Negev, P.O.B 653 Beer-Sheva 8410501, Israel

[7] Tel Aviv University Center for Nanoscience and Nanotechnology, Tel Aviv University, P.O.B 39040, Ramat Aviv 6997801, Israel

### Calculation of capacitance, energy density, and power density

The electrochemical performance of the MXene@VMT microsupercapacitors was evaluated using cyclic voltammetry (CV) and galvanostatic charge–discharge (GCD) measurements in a two-electrode configuration, over a bipolar voltage range of −1.25 to +1.25 V. Areal capacitance was calculated from the integrated area of the CV loop (Eq. 1), where $S_{CV}$ is the enclosed area (A V), ν is the scan rate (V $s^{-1}$), and the factor of two accounts for the forward and reverse scans. Independently, GCD capacitance was determined from the discharge charge $Q_{dis}$ of one complete branch (Eq. 2). In both calculations, ΔV = 2.5 V and A represents the area of a single MXene electrode.

$$C_{A,\mathrm{CV}} = \frac{S_{\mathrm{CV}}}{2\nu A \Delta V} \tag{1}$$

$$C_A = \frac{Q_{\mathrm{dis}}}{A\Delta V} \tag{2}$$

GCD measurements were performed over a bipolar voltage range, sweeping from +1.25 to −1.25 V, which involves discharging to zero potential followed by charging under reversed polarity. Therefore, the full 2.5 V interval does not represent a single energy-delivering discharge. Energy and power densities were calculated from the positive-polarity discharge interval (+1.25 to 0 V), while the negative-polarity response was evaluated separately.

The delivered energy was calculated by numerical integration of the experimental GCD curves over the positive-polarity discharge interval (Eq. 3), where V(t) and I(t) are the measured voltage and current, respectively. This approach accounts for nonlinear discharge behavior and resistive voltage losses.

$$E_{\mathrm{dis}} = \int_{t_1}^{t_2} |V(t)I(t)|\ dt \tag{3}$$

The areal and volumetric energy densities were calculated by normalizing the delivered energy to the area and combined volume of the MXene electrodes, respectively (Eq. 4 and 5). For the e4 device, the area of a single MXene electrode was A = 0.038375 $cm^2$, and the electrode thickness was h = 8 nm. The volumetric energy density was normalized to the combined volume of both electrodes (2Ah = $6.14 \cdot 10^{-8}$ $cm^3$).

$$E_A = \frac{E_{\mathrm{dis}}}{2A} \tag{4}$$

$$E_V = \frac{E_{\mathrm{dis}}}{2A\,h} \tag{5}$$

The average areal and volumetric power densities were calculated by dividing the delivered energy by the discharge duration ($\Delta t_{dis}$), using the same area and volume normalization as for the energy densities (Eq. 6 and 7).

$$P_A = \frac{E_{\mathrm{dis}}}{2A\,\Delta t_{\mathrm{dis}}} \tag{6}$$

$$P_V = \frac{E_{\mathrm{dis}}}{2A\,h\,\Delta t_{\mathrm{dis}}} \tag{7}$$

Energy and power densities were expressed in μWh $cm^{-2}$, mWh $cm^{-3}$, μW $cm^{-2}$, and W $cm^{-3}$.

The Coulombic efficiency and capacitance retention during prolonged GCD cycling were calculated using Eqs. 8 and 9, respectively.

$$\eta_Q = \frac{Q_{\mathrm{dis}}}{Q_{\mathrm{ch}}} \times 100\% \tag{8}$$

$$R_C = \frac{C_{A,N}}{C_{A,0}} \times 100\% \tag{9}$$

Here, $Q_{ch}$ is the charge transferred during charging, while $C_{A,N}$ and $C_{A,0}$ represent the areal capacitances at cycle N and the initial cycle, respectively.

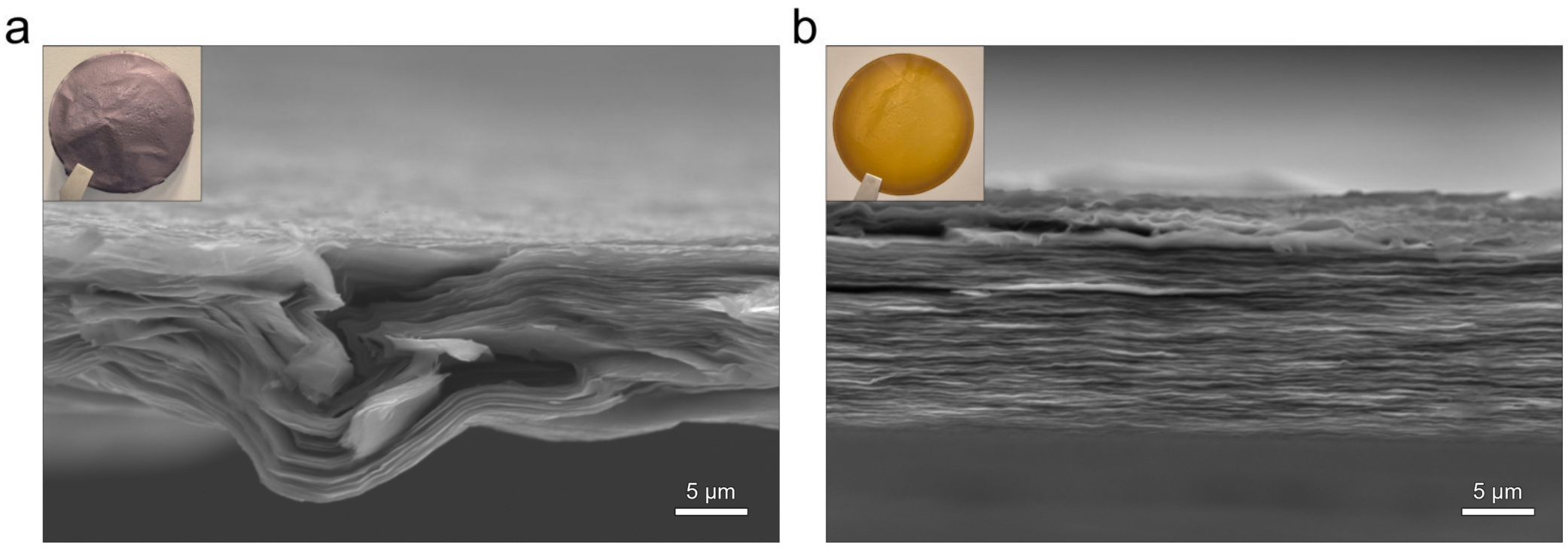

**Fig. S1. MXene and VMT free-standing films.** Cross-sectional SEM images of **a** MXene and **b** vermiculite free-standing films. Insets: optical images of the corresponding free-standing films (4 cm diameter).

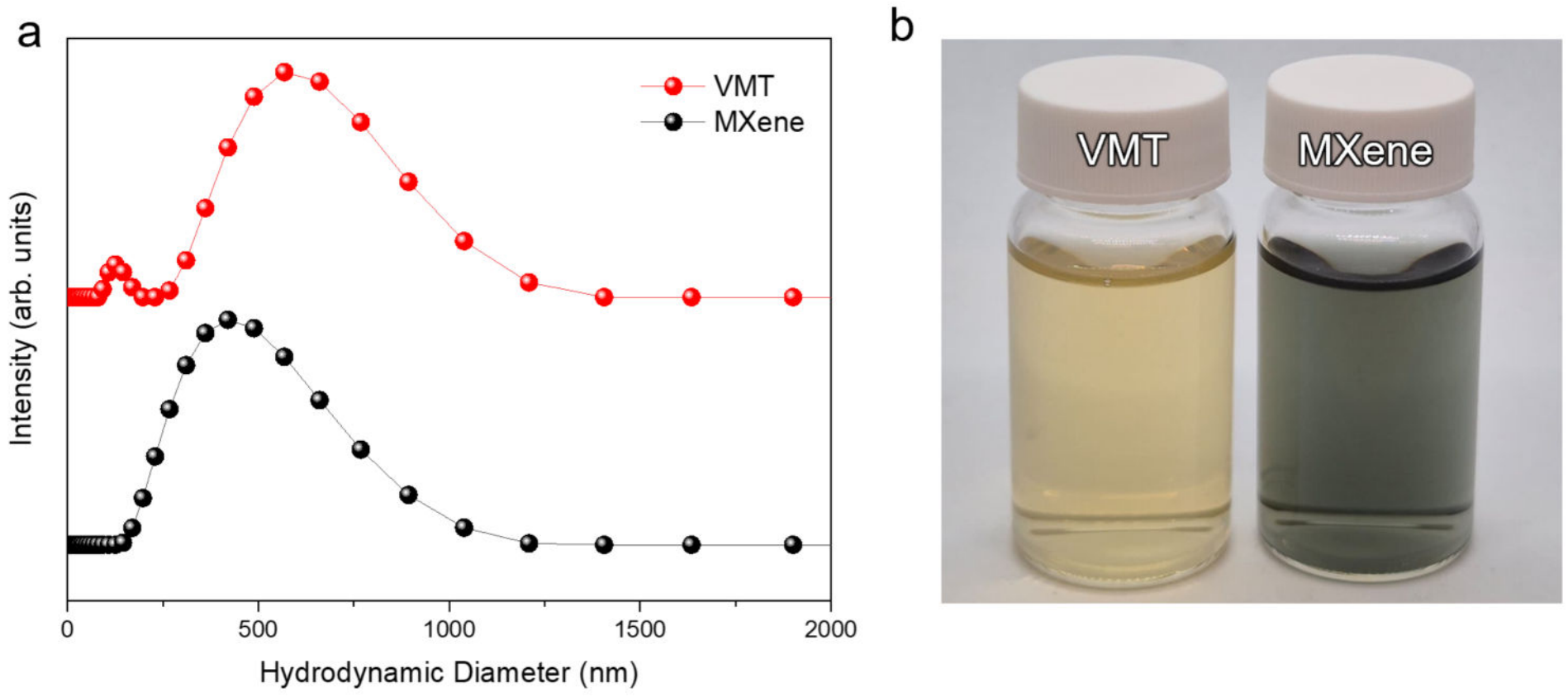

**Fig. S2. Flake size analysis. a** Dynamic light scattering (DLS) of MXene (475 nm) and VMT (615 nm). **b** Photographs of the measured solutions.

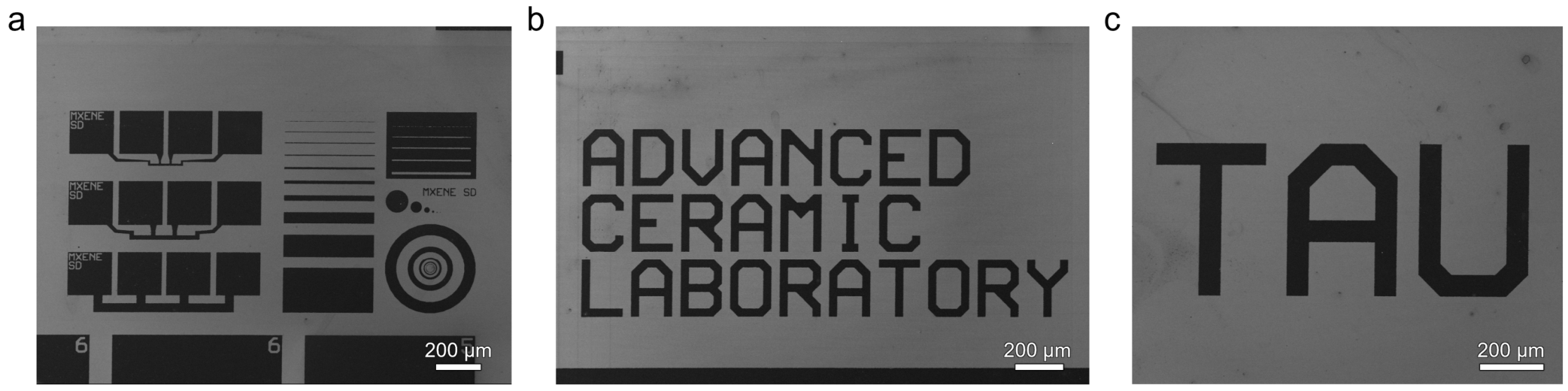

**Fig. S3. MXene deposition monitoring. a–c** SEM images of patterned MXene calibration marks.

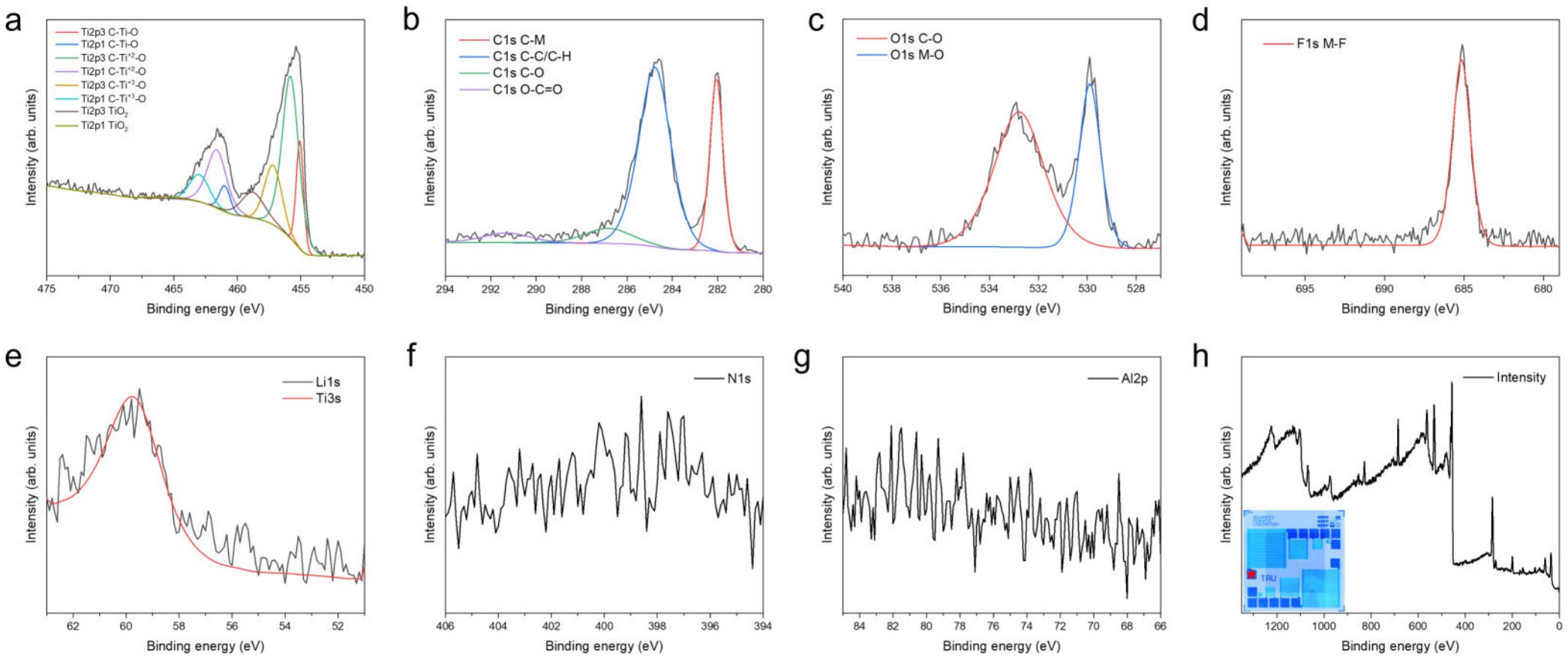

**Fig. S4. XPS analysis of the MXene ped. a** Ti2p, **b** C1s, **c** O1s, **d** F1s, **e** Li1s, **f** N1s, **g** Al2p and **h** survey spectrum (inset shows optical image of the chip with the analyzed MXene ped marked with red star).

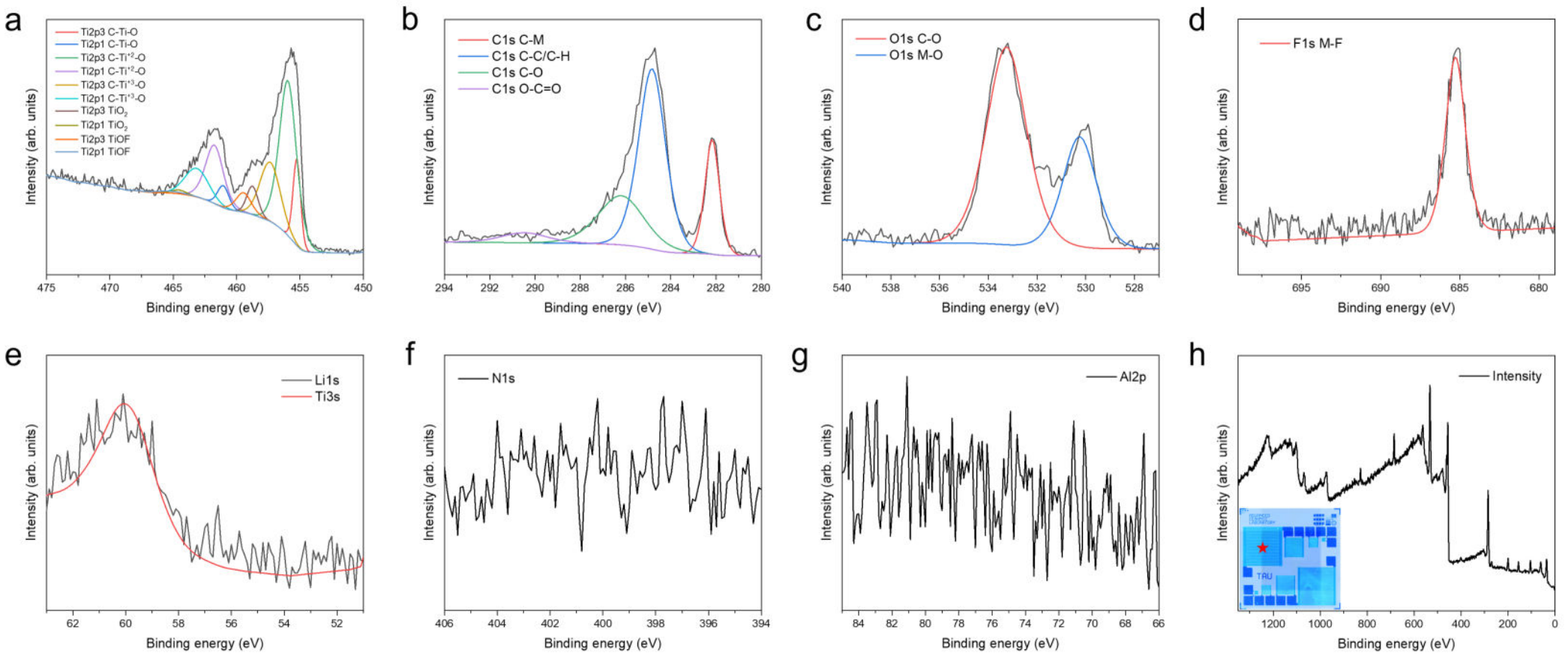

**Fig. S5. XPS analysis of the MXene electrodes. a** Ti 2p, **b** C 1s, **c** O 1s, **d** F 1s, **e** Li 1s, **f** N 1s, **g** Al 2p and **h** survey spectrum (inset shows optical image of the chip with the analyzed MXene electrode marked with red star).

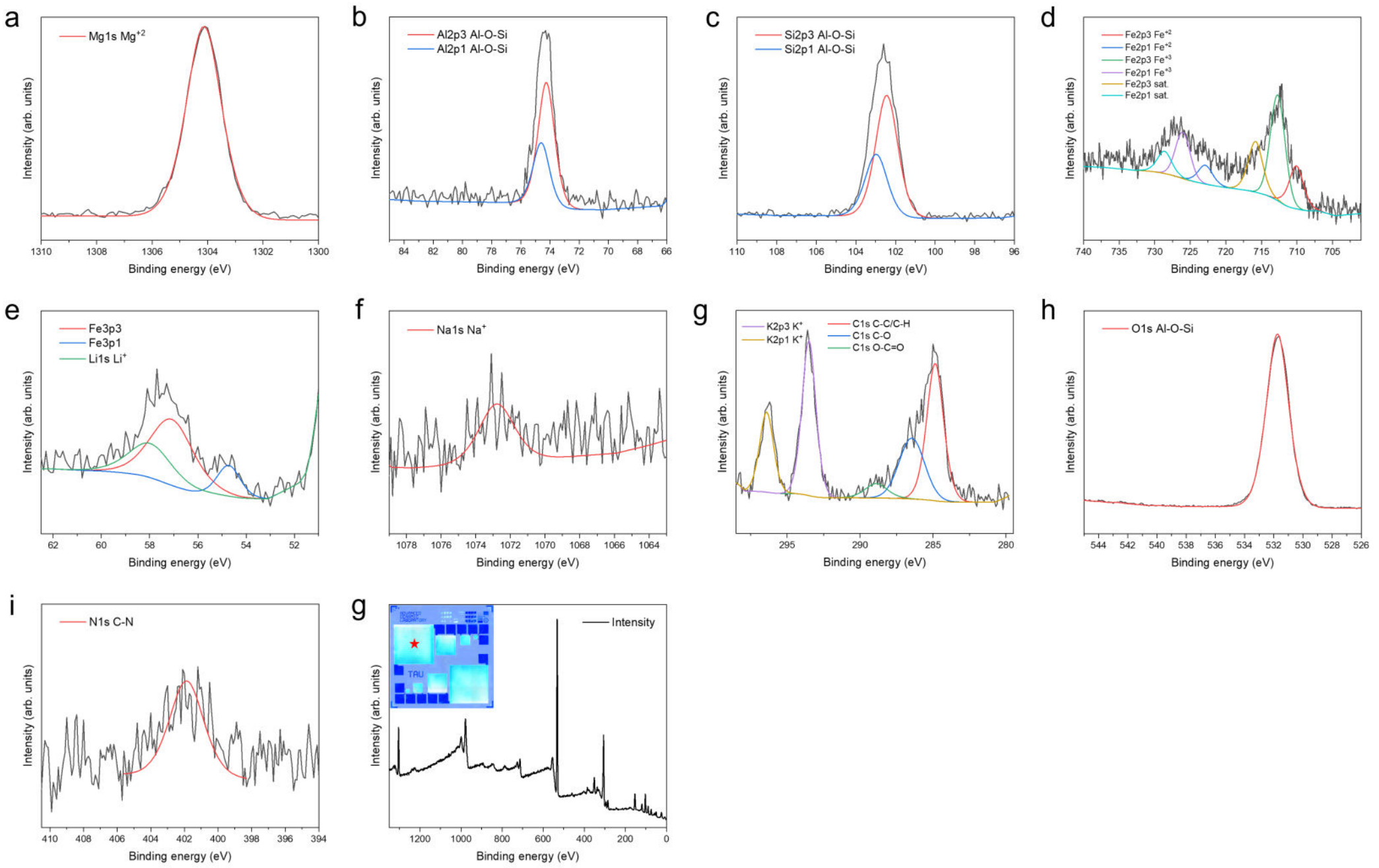

**Fig. S6. XPS analysis of the VMT electrolyte on MXene electrodes. a** Mg 1s, **b** Al 2p, **c** Si 2p, **d** Fe 2p, **e** Li 1s, **f** Na 1s, **g** C 1s and K 2p, **h** O 1s, **i** N 1s and **g** survey spectrum (inset shows optical image of the chip with the analyzed VMT@MXene microsupercapacitor marked with red star).

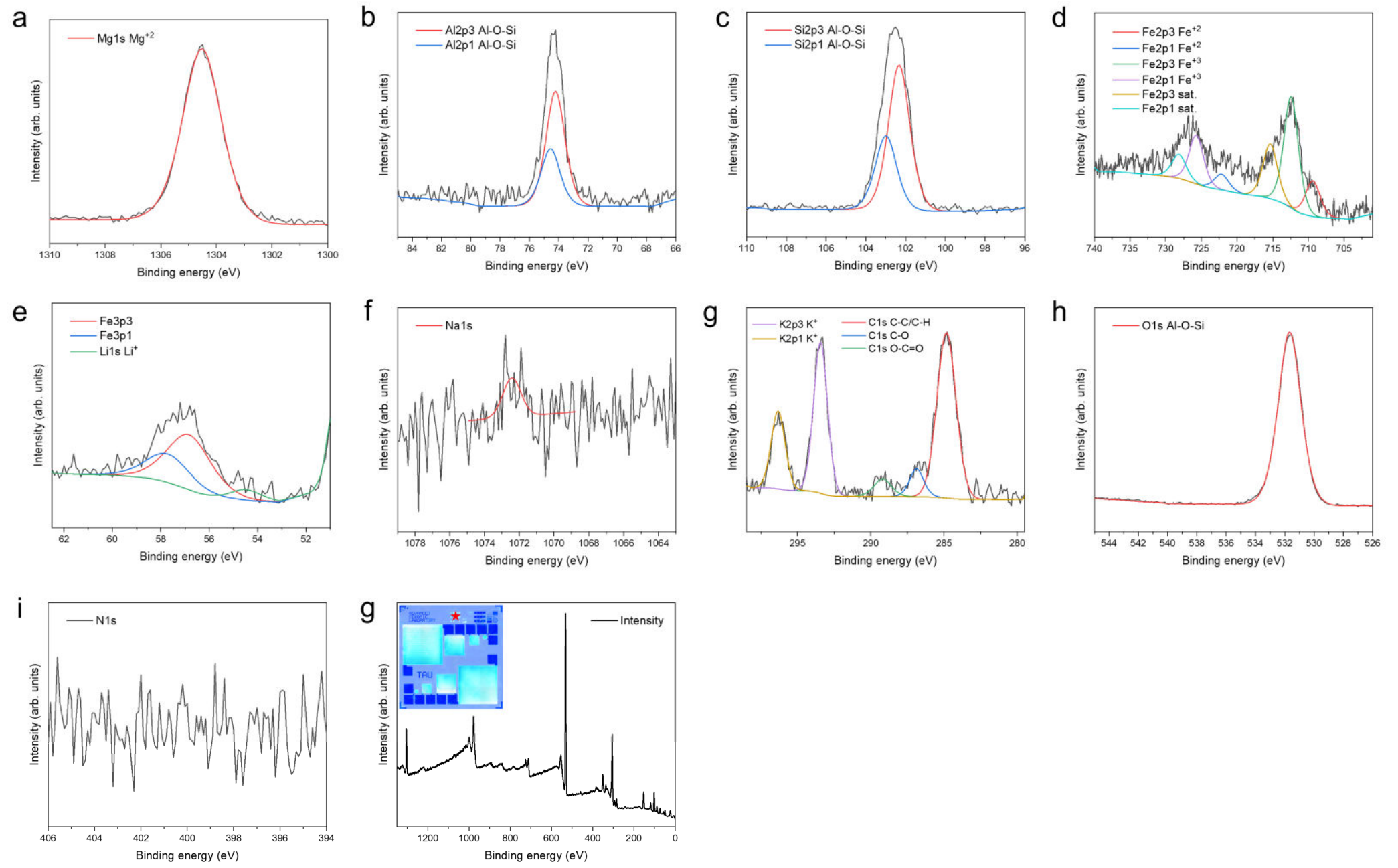

**Fig. S7. XPS analysis of the VMT ped. a** Mg 1s, **b** Al 2p, **c** Si 2p, **d** Fe 2p, **e** Li 1s, **f** Na 1s, **g** C 1s and K 2p, **h** O 1s, **i** N 1s and **g** survey spectrum (inset shows optical image of the chip with the analyzed VMT ped marked with red star).

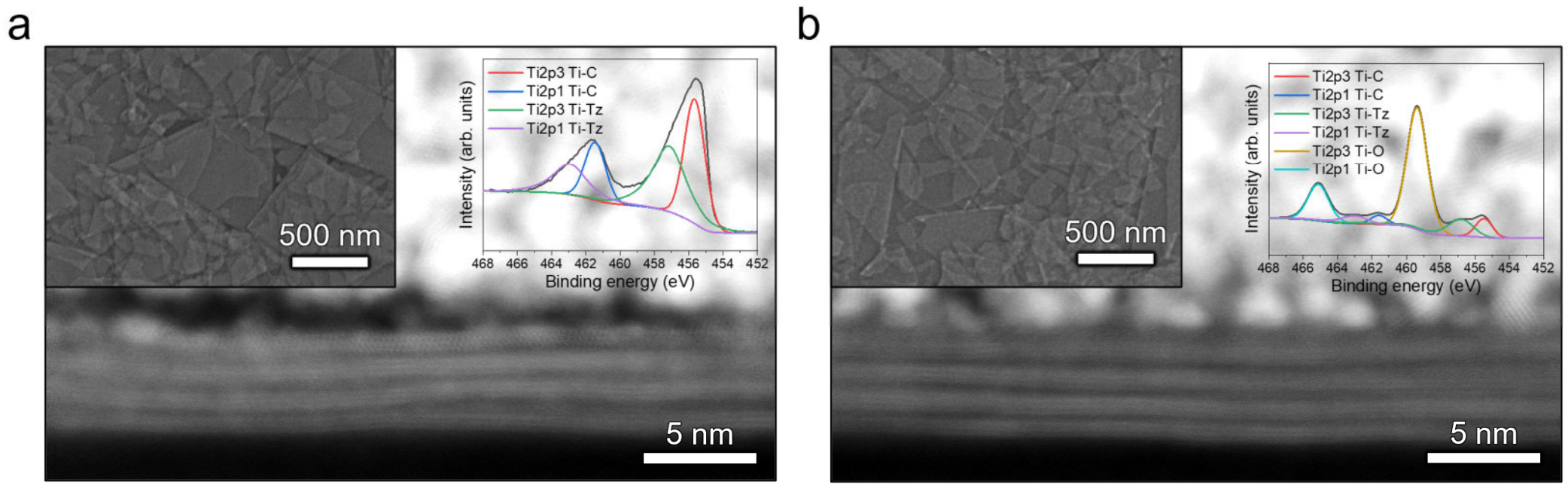

**Fig. S8. Effect of UV ozone treatment on MXene morphology.** Cross-sectional STEM images of **a** pristine deposited MXene and **b** MXene after 20 min of UV–ozone exposure. Insets: top-view SEM images (left) and Ti 2p XPS peak fitting (right).

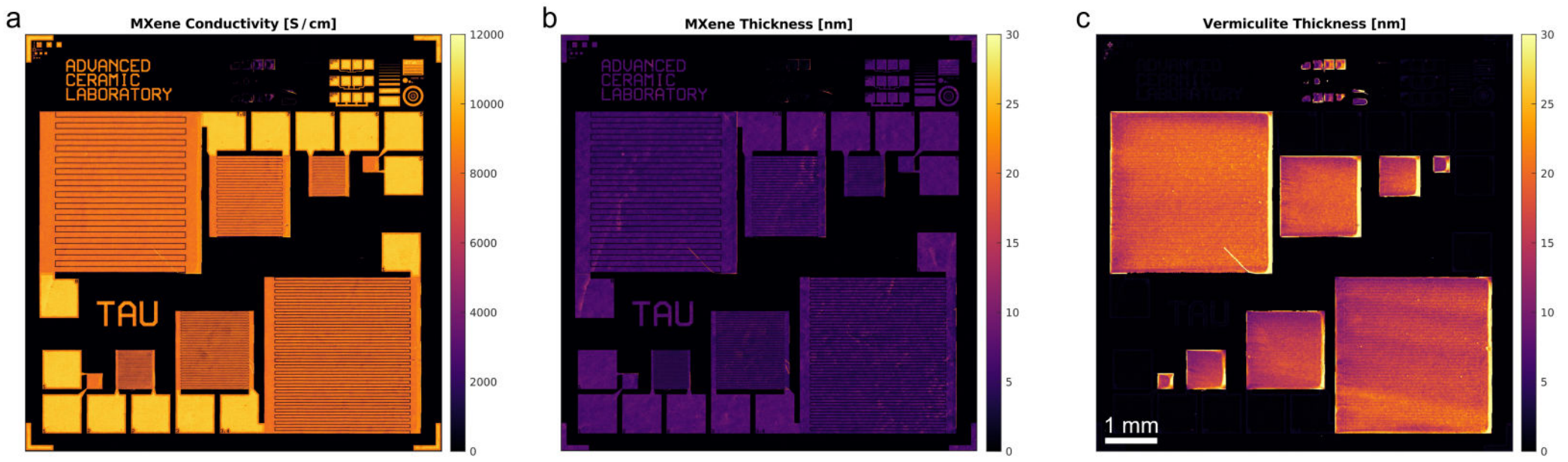

**Fig. S9. Micro-imaging ellipsometry of completed chip at higher magnification.** **a** Conductivity image of the MXene electrodes. **b** Thickness image of the MXene film. **c** Thickness image of the VMT film.

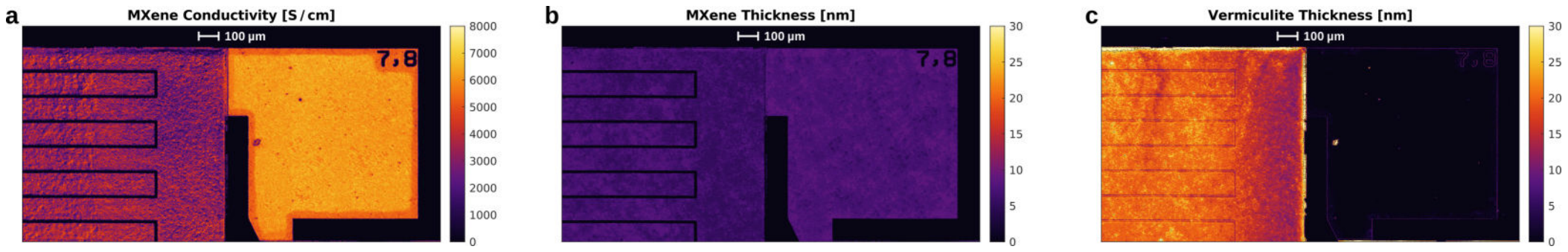

**Fig. S10. Micro-imaging ellipsometry of the MXene@VMT devices at higher magnification.** **a** Conductivity image of the MXene electrodes. **b** Thickness image of the MXene film. **c** Thickness image of the VMT film.

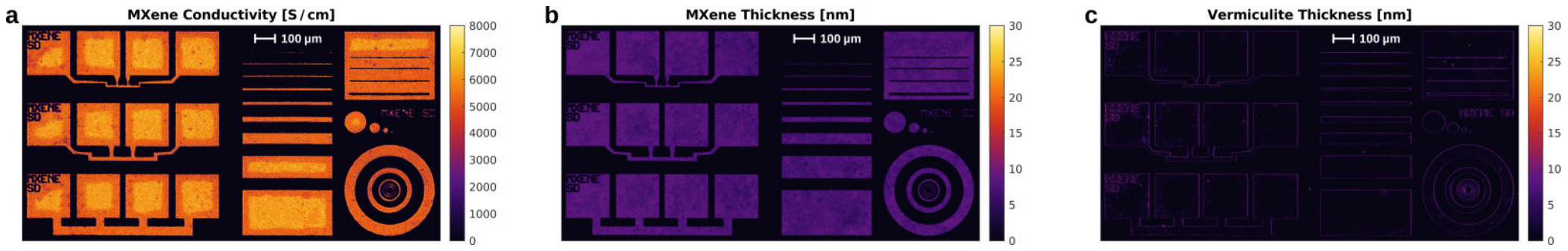

**Fig. S11. Micro-imaging ellipsometry of the MXene calibration pattern.** **a** Conductivity image of the MXene pattern. **b** Thickness image of the MXene film. **c** Apparent VMT thickness image showing the absence of VMT in untreated MXene areas.

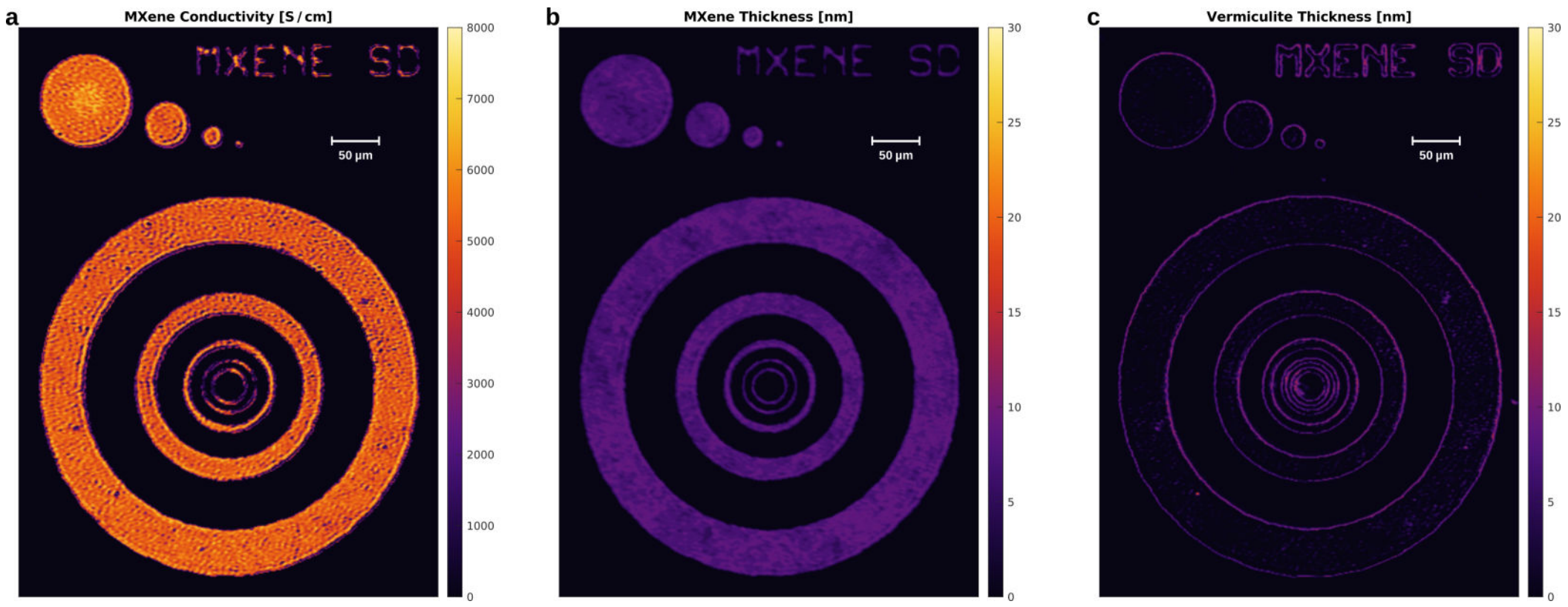

**Fig. S12. Micro-imaging ellipsometry of the MXene calibration pattern. a** Conductivity image of the MXene pattern. **b** Thickness image of the MXene film. **c** Apparent VMT thickness image showing the absence of VMT in untreated MXene areas.

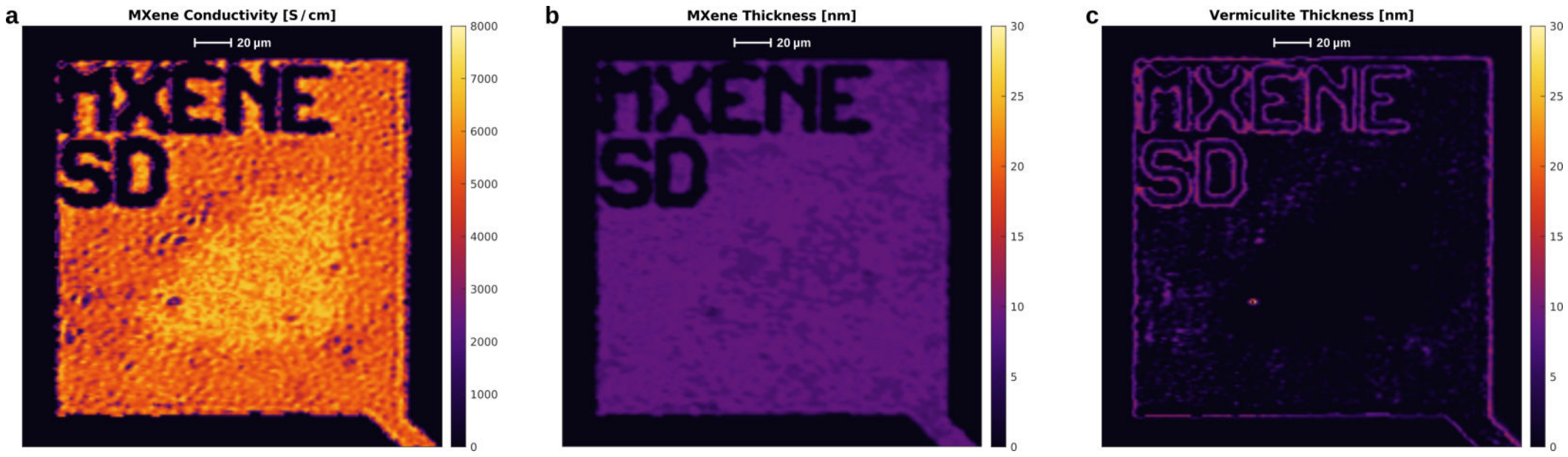

**Fig. S13. Micro-imaging ellipsometry of an MXene pad. a** Conductivity image of the MXene pad. **b** Thickness image of the MXene film. **c** Apparent VMT thickness image showing the absence of VMT in untreated MXene areas.

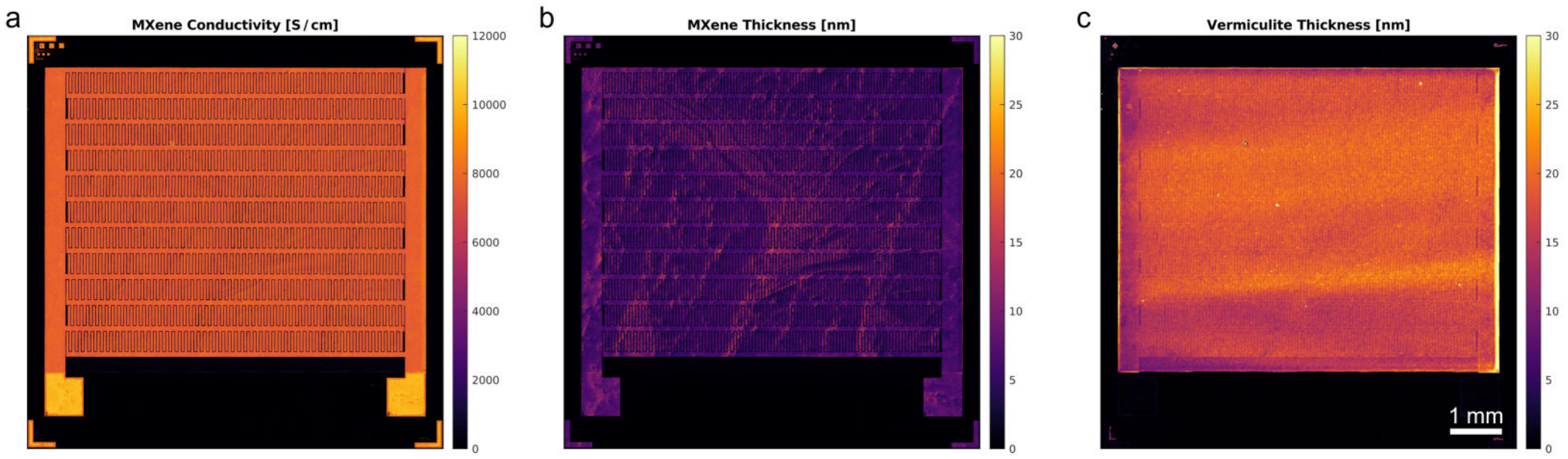

**Fig. S14. Micro-imaging ellipsometry of the large-area MXene@VMT device. a** Conductivity image of the MXene electrode. **b** Thickness image of the MXene film. **c** Thickness image of the VMT film.

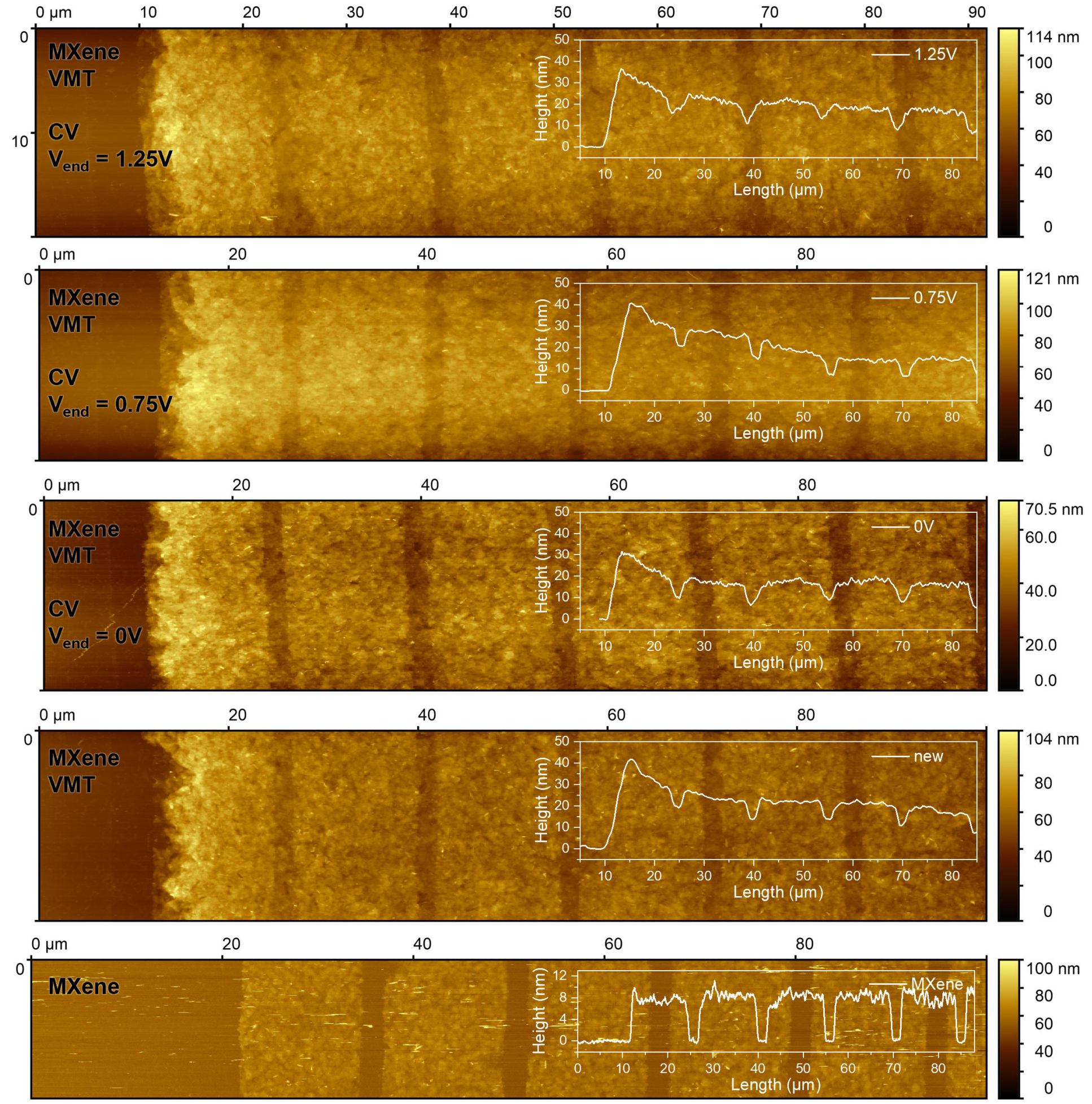

**Fig. S15. AFM analysis of the MXene@VMT microsupercapacitor.** AFM scans of a bare MXene electrode and of the same electrode with VMT electrolyte (e2) after CV, stopped at different potentials, with an inset showing the average height profile.

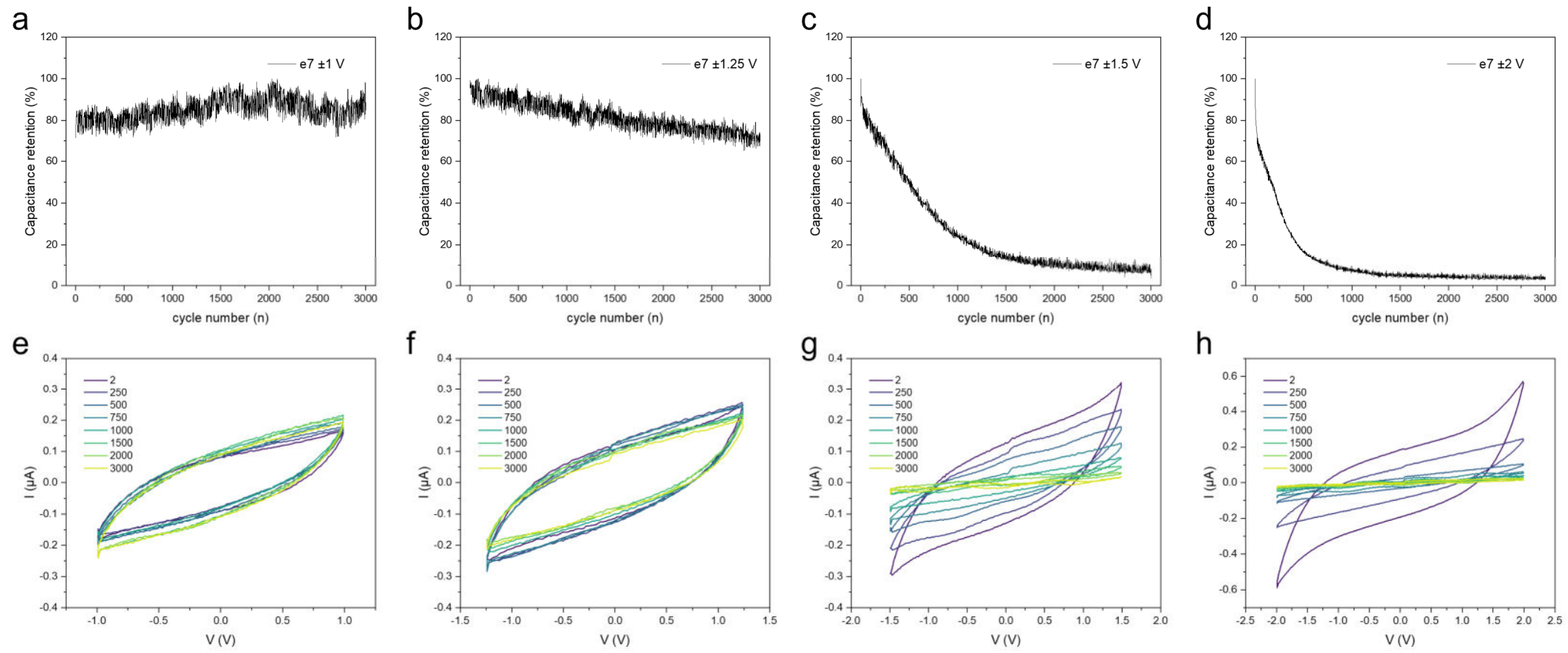

**Fig. S16. MXene@VMT operation window.** Capacitance retention over 3000 cycles and CV curves at (**a,e**) ±1 V, (**b,f**) ±1.25 V, (**c,g**) ±1.5 V, and (**d,h**) ±2 V (scan rate: 1 V $s^{-1}$).

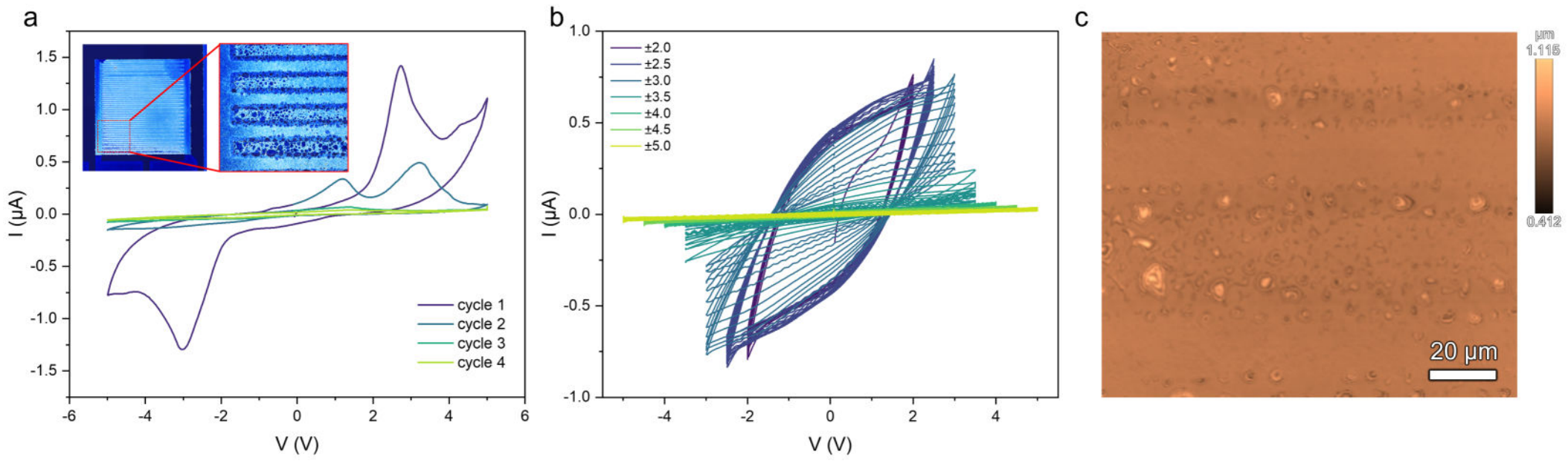

**Fig. S17. Oxidation behavior of the MXene@VMT microsupercapacitor. a** CV at a potential window of ±5 V (50 mVs$^{-1}$), inset: optical microscopy image of MXene@VMT after oxidation. **b** CV at different potential windows (200 mVs$^{-1}$). **c** Confocal height map of the oxidized MXene electrode and bubble formation.

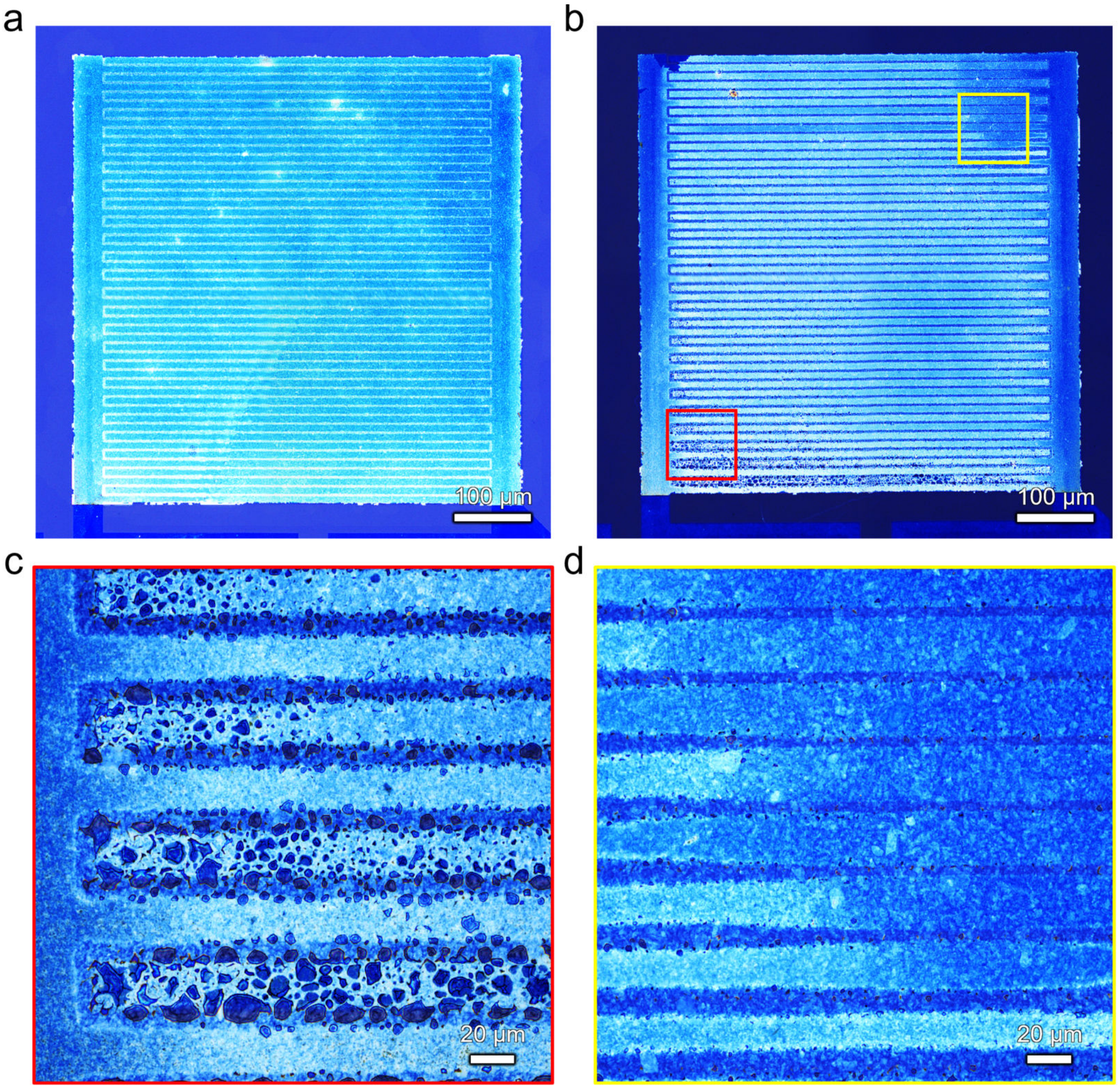

**Fig. S18. Optical microscopy images of defects in the MXene@VMT microsupercapacitor after oxidation.** Optical microscopy images of new (**a**) and oxidized (±5 V cycled) (**b**) MXene@VMT. **c** Increased magnification showing bubble formation underneath the VMT electrolyte. **d** Disconnected electrodes after oxidation.

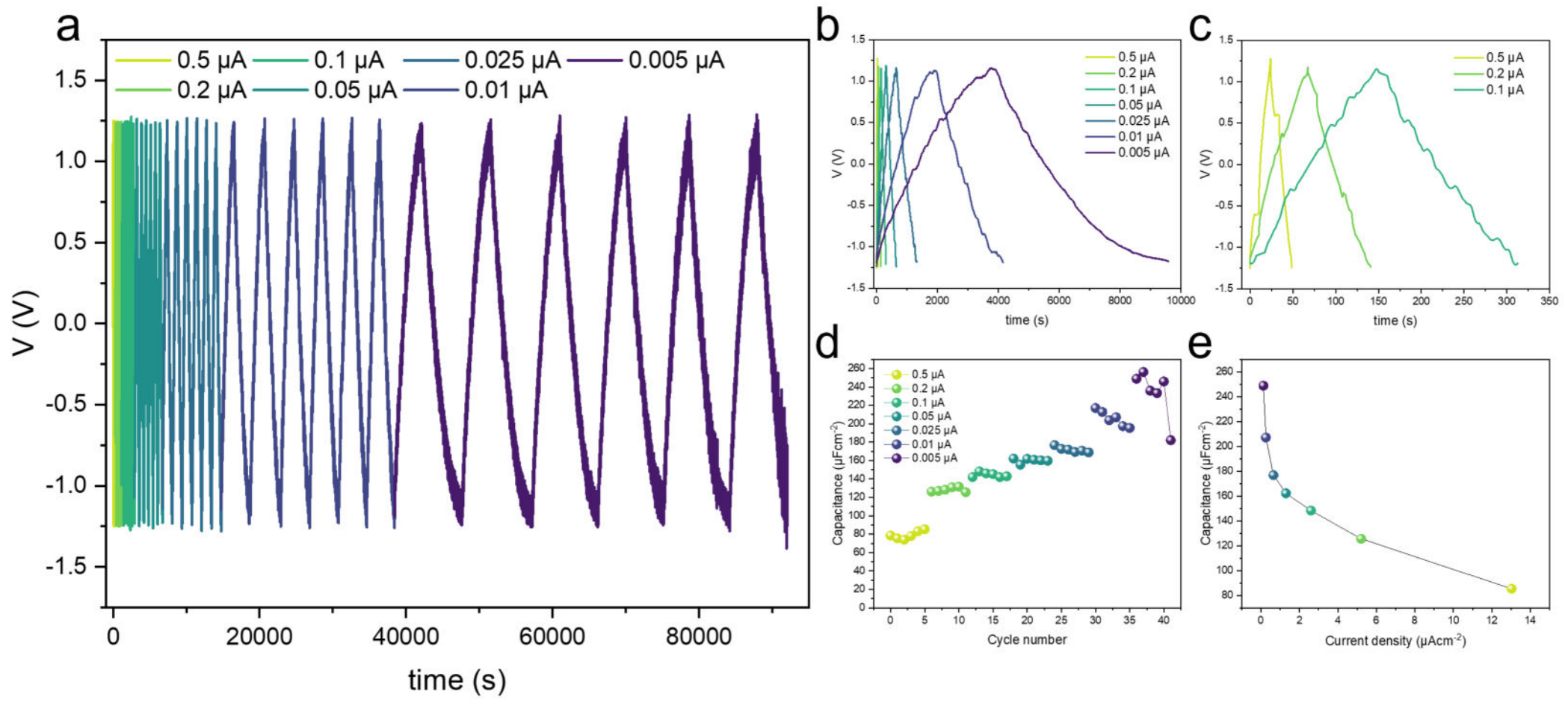

**Fig. S19. Galvanostatic characterization of the MXene@VMT microsupercapacitor (e4 device)**. **a** GCD response at different applied currents. **b,c** Representative charge–discharge curves showing current dependent behavior. **d** Capacitance extracted from the GCD profiles. **e** Capacitance as a function of applied current, reflecting rate capability.

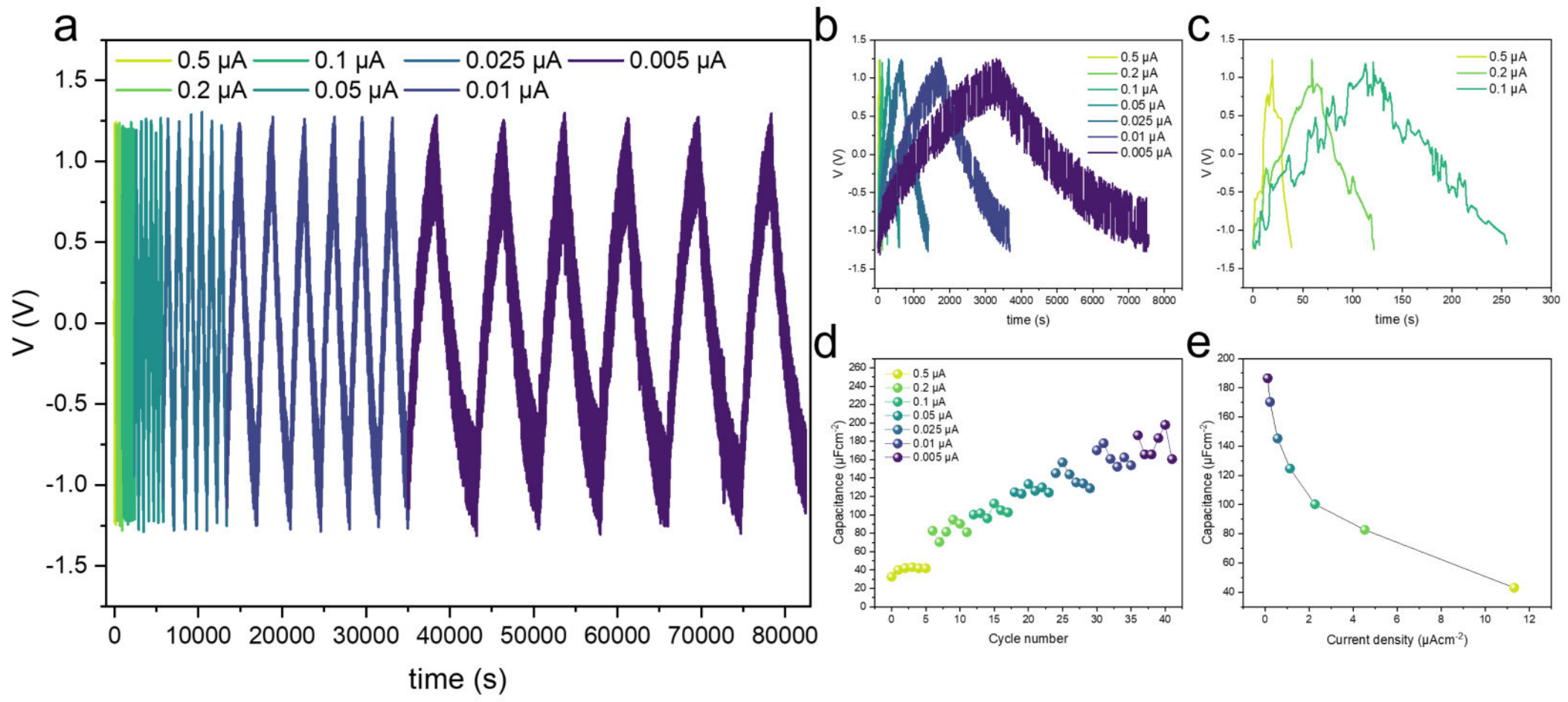

**Fig. S20. Galvanostatic characterization of the MXene@VMT microsupercapacitor (e8 device)**. **a** GCD response at different applied currents. **b,c** Representative charge–discharge curves showing current dependent behavior. **d** Capacitance extracted from the GCD profiles. **e** Capacitance as a function of applied current, reflecting rate capability.

**Table S1.** Comparison of the MXene@VMT microsupercapacitor with representative on-chip microsupercapacitors.

| **Work** | **Electrode material** | **Electrolyte material** | **Electrode Thickness** | **Electrolyte Thickness** | **Voltage** | **Areal Capacitance (mF cm$^{-2}$)** | **Areal Energy (μWh cm$^{-2}$)** | **Volumetric* Energy (mWh cm$^{-3}$)** | **Power (W cm$^{-3}$)** | **Reference** |
|---|---|---|---|---|---|---|---|---|---|---|
| **This work** | MXene | Solid (2D VMT) | ~8 nm | ~20 nm | 1.25 V | ~0.24 | ~0.012 | ~15.5 | ~1.4 | This work |
| **Jiang et al.** (*Adv. Energy Mater.*, 2019) | Au/MXene | (PVA)/$H_3PO_4$ gel | ~100 nm | NA | 0.6 V | ~0.1-0.5 | ~0.03 | ~2-3 | NA | 1 |
| **Hota et al.** (*Adv. Mater.*, 2019) | $RuO_2$ | (PVA)/$H_2SO_4$ gel | ~125-750 nm | NA | 1 V | 3-5 | ~0.48 | ~19 | ~2 | 2 |
| **Lee et al.** *(Sci. Adv., 2020)* | nAC/CMC/Super-P/Ti/Au | [EMIM][TFSI]/thiol–ene gel | ~2.2–2.3 μm | NA | 1.2 V | ~0.128 | ~0.026 | ~0.11 | NA | 3 |
| **Wang et al.** (*Nat. Commun.*, 2024) | Au/MXene | Ionic liquid ($EMImBF_4$) | ~250 nm | NA | ~1 V | ~2.3 | ~0.42 | ~15–16 | NA | 4 |

* Volumetric values were calculated based on the active electrode volume to enable direct comparison with previously reported microsupercapacitors.

1. **On-Chip MXene Microsupercapacitors for AC-Line Filtering Applications** (*Adv. Energy Mater.*, 2019)
2. **Integration of Electrochemical Microsupercapacitors with Thin Film Electronics for On-Chip Energy Storage** (*Adv. Mater.*, 2019)
3. **Ultrahigh areal number density solid-state on-chip microsupercapacitors via electrohydrodynamic jet printing** (*Sci. Adv.*, 2020)
4. **Monolithically integrated micro-supercapacitors with high areal number density produced by surface adhesive-directed electrolyte assembly** (*Nat. Commun.*, 2024)

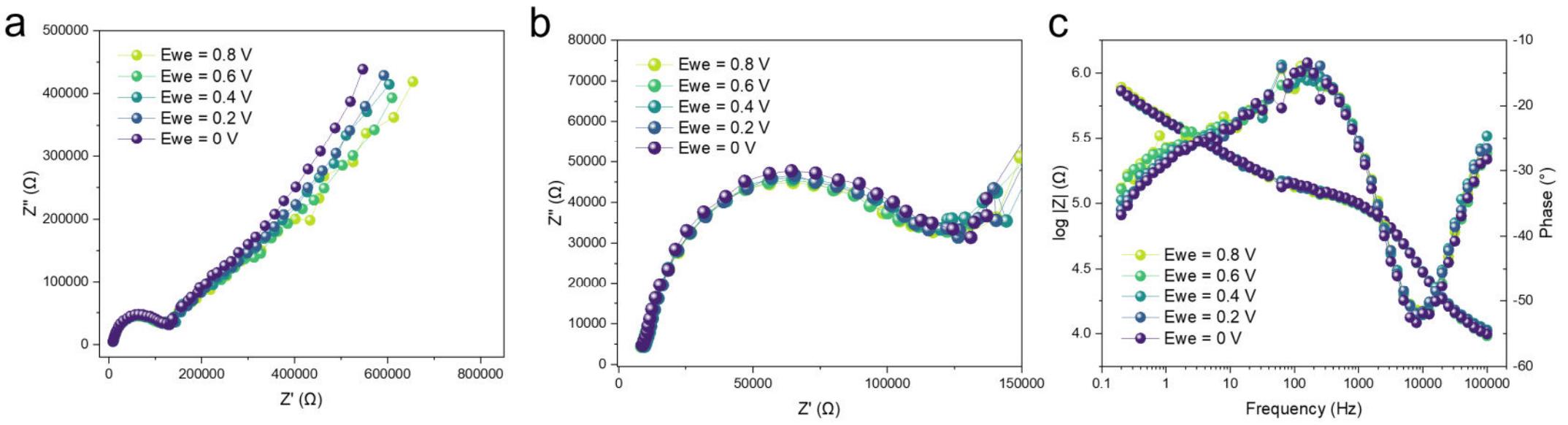

**Fig. S21. MXene@VMT electrochemical impedance spectroscopy. a** Nyquist plots at different bias potentials. **b** Magnified Nyquist plots at high frequencies. **c** Bode plots at different bias potentials.

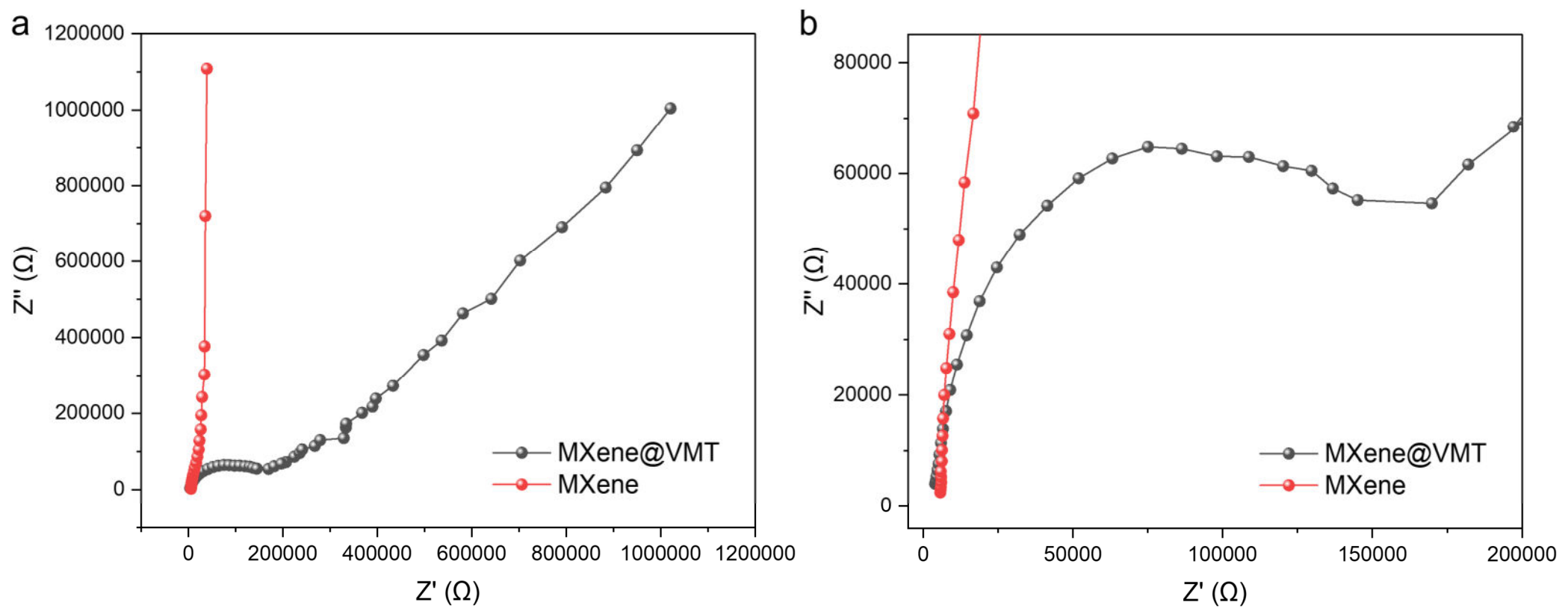

**Fig. S22. Electrochemical impedance spectroscopy of MXene electrodes with and without the VMT electrolyte**. (a,b) Nyquist plots of e8 MXene electrodes measured in the absence and presence of the VMT electrolyte. Panel (b) shows a magnified view of the low-impedance region.

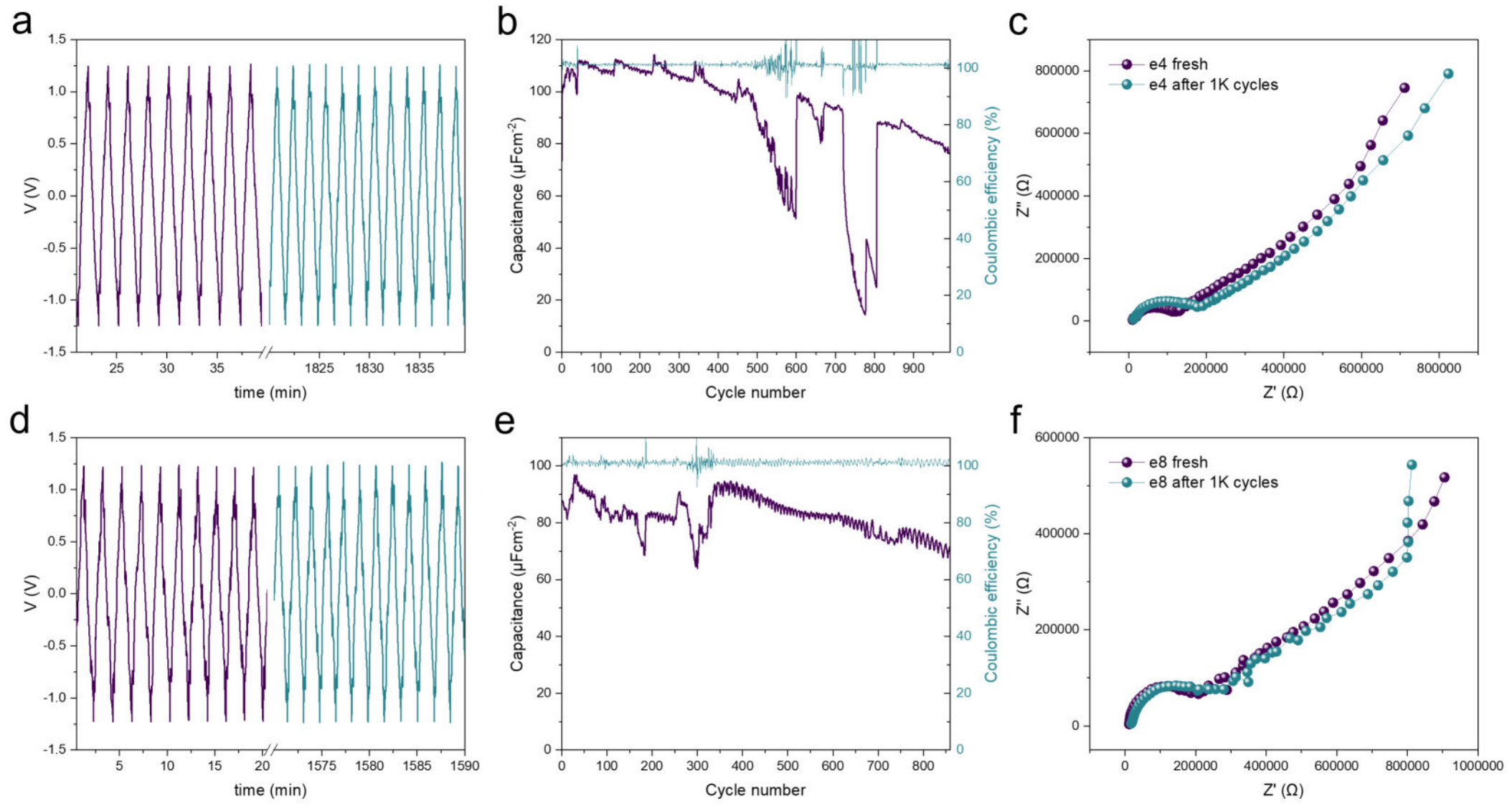

**Fig. S23. Stability test of the MXene@VMT microsupercapacitor.** GCD over 900 cycles (a,d), capacitance (b,e), and EIS before and after cycling (c,f) for e4 and e8, respectively. All measurements were performed at 5 μA $cm^{-2}$.

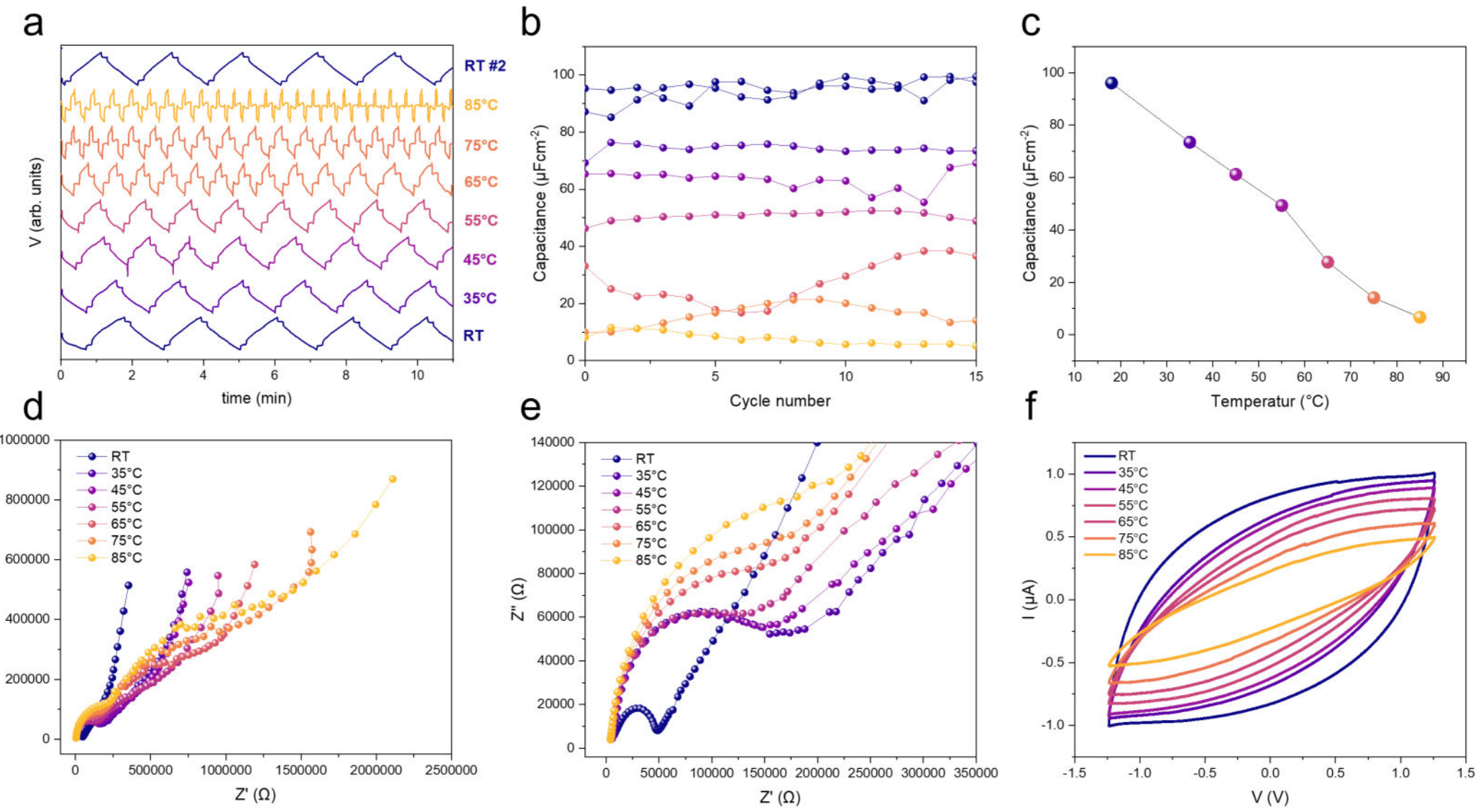

**Fig. S24. Temperature-dependent electrochemical characterization of the MXene@VMT microsupercapacitor. a** GCD response at different temperatures obtained at a current density of 5 µA $cm^{-2}$. **b** Capacitance extracted from the GCD profiles. **c** Capacitance as a function of temperature. **d** EIS spectra at different temperatures. **e** Magnified high-frequency region of the EIS spectra. **f** CV curves at different temperatures at scan rates of 200 $mVs^{-1}$.

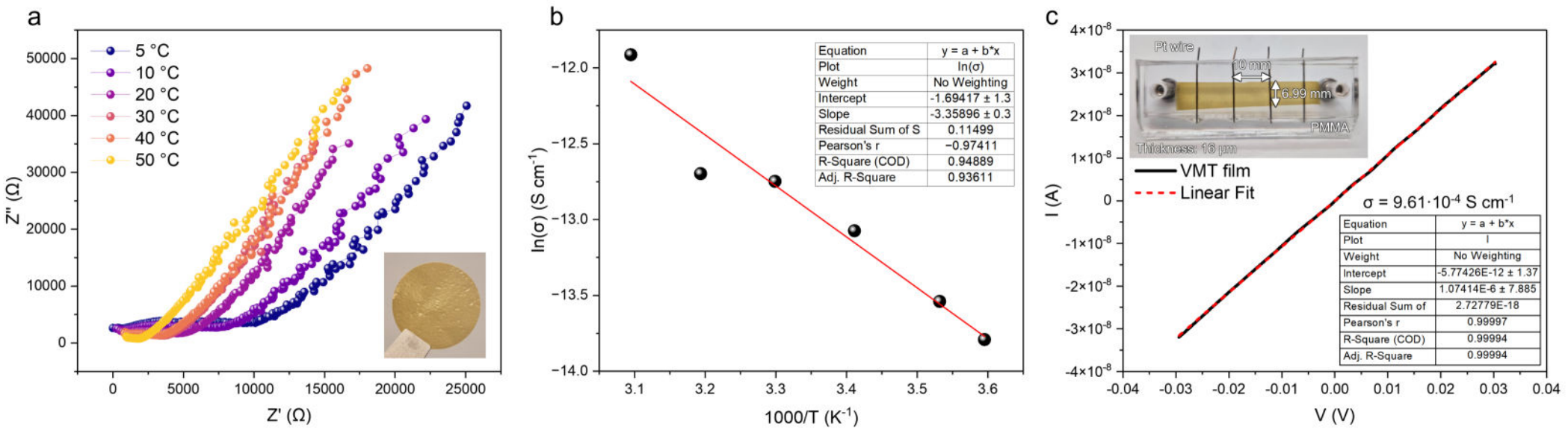

**Fig. S25. Ionic transport characterization of free-standing VMT films. a** Temperature-dependent electrochemical impedance spectra recorded between glassy carbon electrodes under controlled relative humidity. Inset: photograph of the free-standing VMT film. **b** Arrhenius plot of the out-of-plane ionic conductivity, showing thermally activated transport with an apparent activation energy of ~0.29 eV. **c** In-plane four-terminal IV measurement of a free-standing VMT film, showing a conductivity of $9.61 \cdot 10^{-4}$ S $cm^{-1}$. Inset: photograph of the measurement setup.

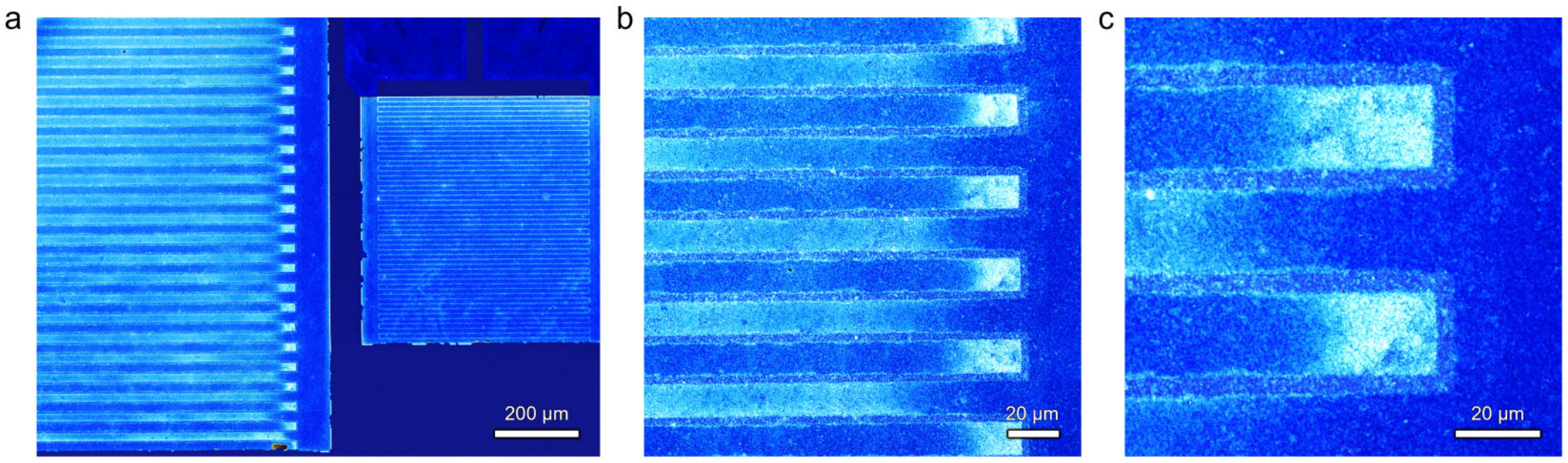

**Fig. S26. Optical microscopy images of the used MXene@VMT microsupercapacitor. a–c** Various magnifications of a representative microsupercapacitor after 1000 cycles.

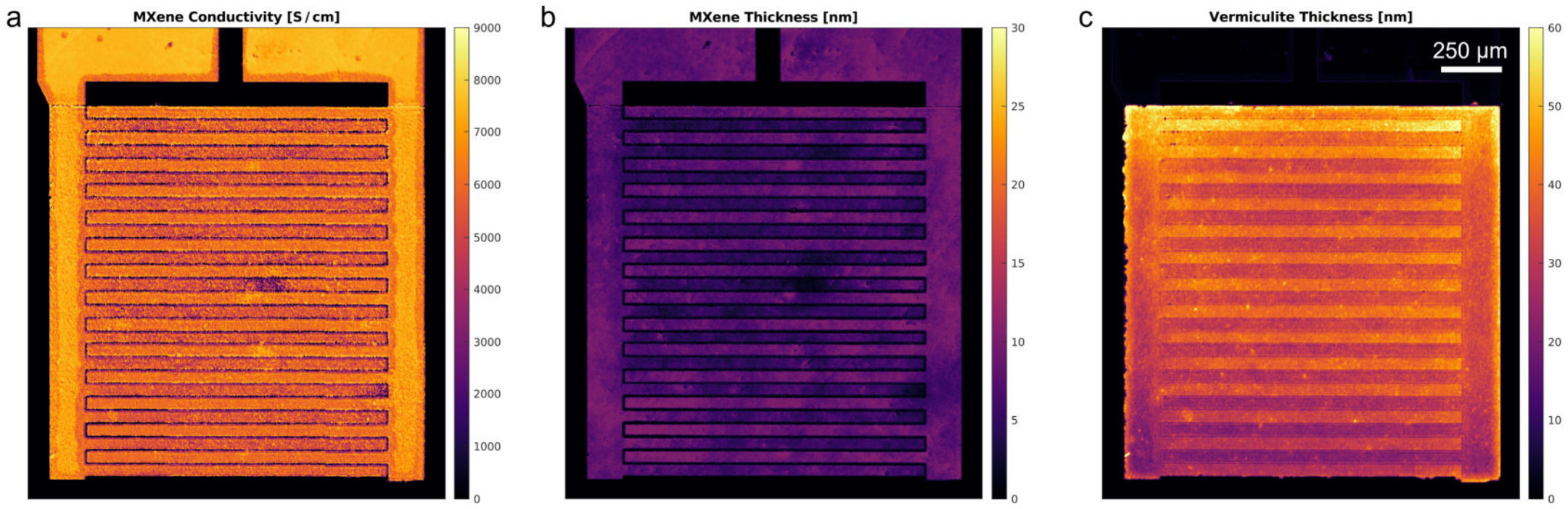

**Fig. S27. Micro-imaging ellipsometry of the MXene@VMT device after oxidation at 5 V. a** conductivity image of the MXene electrodes. **b** thickness image of the MXene film. **c** thickness image of the VMT film.